\documentclass[12pt,a4paper]{article}

\usepackage[english]{babel}
\usepackage{fancyhdr}
\usepackage{amsmath}
\usepackage{amssymb}
\usepackage{amsfonts}
\usepackage{psfrag}
\usepackage{inputenc}
\usepackage{graphicx,wrapfig}
\usepackage[bf,footnotesize]{caption2}
\usepackage{cite}

\usepackage{bm}
\usepackage[dvipsnames]{xcolor}
\usepackage[colorlinks=True, citecolor=blue, linkcolor=blue, urlcolor=blue,linktocpage]{hyperref}
\usepackage[sort&compress,numbers, merge]{natbib}

\newcommand{\mum}{\,\mu\mathrm{m}}
\newcommand{\km}{\,\mathrm{km}}

\newcommand{\vev}[1]{\langle #1 \rangle}

\newcommand{\UV}{\mathrm{UV}}

\newcommand{\Lag}{\mathcal{L}}
\newcommand{\amp}{\mathcal{M}}

\newcommand{\beq}{\begin{equation}}
\newcommand{\eeq}{\end{equation}}
\newcommand{\bea}{\begin{eqnarray}}
\newcommand{\eea}{\end{eqnarray}}

\DeclareRobustCommand{\Sec}[1]{Sec.~\ref{#1}}

\DeclareRobustCommand{\App}[1]{App.~\ref{#1}}

\DeclareRobustCommand{\Fig}[1]{Fig.~\ref{#1}}

\DeclareRobustCommand{\Eq}[1]{Eq.~(\ref{#1})}

\newcommand{\mpl}{M_{\rm Pl}}
\newcommand{\invkm}{\,\mathrm{km}^{-1}}

\newcommand{\GR}{\mathrm{GR}}
\newcommand{\BGR}{\mathrm{BGR}}
\newcommand{\GB}{\mathrm{GB}}

\newcommand{\CS}{\mathrm{CS}}
\newcommand{\GBsc}{\mathrm{GBization}}

\newcommand{\La}{\Lambda_\alpha}
\newcommand{\Lat}{\Lambda_{\tilde{\alpha}}}

\begin{document}

\begin{titlepage}
	\begin{center}
	
		\vspace{2.0cm}
		{\Large\bf  
			Causality Constraints on Black Holes beyond GR}
			
		\vspace{1.0cm}
		\renewcommand{\thefootnote}{\fnsymbol{footnote}}
		{\small \bf 
			Francesco Serra$^{a,b}$,
			Javi Serra$^{c}$,
			Enrico Trincherini$^{a,b}$ and
			Leonardo G.~Trombetta$^{d}$\, \footnote{E-mail:
				\href{francesco.serra@sns.it}{francesco.serra@sns.it},
				\href{mailto:javi.serra@tum.de}{javi.serra@tum.de},
				\href{enrico.trincherini@sns.it}{enrico.trincherini@sns.it},
				\href{mailto:trombetta@fzu.cz}{trombetta@fzu.cz}}
		}
		
		\vspace{0.7cm}
		{\it\footnotesize
			${}^a$Scuola Normale Superiore, Piazza dei Cavalieri 7, 56126, Pisa, Italy\\
			${}^b$Istituto Nazionale di Fisica Nucleare (INFN) - Largo B.~Pontecorvo, 3, 56127 Pisa, Italy\\
			${}^c$Technische Universit\"{a}t M\"{u}nchen, Physik-Department, 85748 Garching, Germany\\
			${}^d$CEICO, Institute of Physics of the Czech Academy of Sciences, Na Slovance 1999/2, 182 21, Prague 8, Czechia
		}
		
		\vspace{0.9cm}
		\abstract{\noindent We derive causality constraints on the simplest scalar-tensor theories in which black holes differ from what General Relativity 
		predicts, a scalar coupled to the Gauss-Bonnet or the Chern-Simons terms. Demanding that time advances are unobservable within the regime of validity of these effective field theories, we find their cutoff must be parametrically of the same size as the inverse Schwarzschild radius of the black holes for which the non-standard effects are of order one. For astrophysical black holes within the range of current gravitational wave detectors, this means a cutoff length at the km. 
		We further explore the leading additional higher-dimensional operators potentially associated with the scale of UV completion and discuss their phenomenological implications for gravitational wave science.
		}
	\end{center}
\end{titlepage}
\renewcommand{\thefootnote}{\arabic{footnote}}
\setcounter{footnote}{0}


\section{Introduction}
\label{intro}

The detection of gravitational waves from black hole mergers has opened a new window into the nature of gravitational interactions. In particular, the possibility to study gravity in the strong field regime for the first time has motivated a surge of interest in field theories that allow for black hole solutions different from the ones predicted by General Relativity (GR).

In the absence of a compelling guiding principle, the intrinsic complexity of the merger process has encouraged the study of simple models where deviations from GR could be order one.
This is the case of scalar-tensor theories featuring the lowest-dimensional non-minimal couplings of a scalar field to gravity, capable of sourcing detectable scalar hair around black holes: a massless (shift-symmetric) scalar 
coupled to the Gauss-Bonnet (GB) invariant \cite{Sotiriou:2013qea,Sotiriou:2014pfa} or to the Chern-Simons (CS) term, a.k.a.~Pontryagin invariant, \cite{Yunes:2009hc},
\beq
	S = \int d^4 x \sqrt{-g} \left( \frac{\mpl^2}{2} R - \frac{1}{2} (\nabla_\mu \phi)^2 + \mpl \alpha \phi \mathcal{R}_{\GB}^2 + \mpl \tilde{\alpha} \phi R_{\mu\nu\rho\sigma} \tilde R^{\mu\nu\rho\sigma} \right)\, ,
\label{eq:Shair}
\eeq
where $\mathcal{R}_{\GB}^2 \equiv R^{\mu\nu\rho\sigma}R_{\mu\nu\rho\sigma}-4R^{\mu\nu}R_{\mu\nu}+R^2$ and $\tilde R^{\mu\nu\rho\sigma} = \tfrac{1}{2} \epsilon^{\mu\nu \alpha \beta} R_{\alpha \beta}^{\quad \rho \sigma}$. $\alpha$ and $\tilde{\alpha}$ are the length-scales (squared) parametrizing the strength of the non-minimal scalar couplings.
For shift-symmetric scalar theories a no-hair theorem \cite{Hui:2012qt} basically selects these two interactions as the only ones leading to black hole hair \cite{Creminelli:2020lxn}. See \cite{Herdeiro:2015waa} for other types of hair in black hole geometries.

While very interesting from the phenomenological point of view, it is crucial to understand how much one can learn about the fundamental properties of gravity via the study of these models in the context of gravitational wave observations.
To answer this important question, the very first step is to assess the consistency of such extensions of GR with what we already know: that GR provides a good description of gravitational interactions down to $\mu$m scales \cite{Lee:2020zjt} and, at the most basic level, that the principles of unitarity, locality, and causality hold there.

Based on simple causality arguments we will find that the cutoff of the effective field theory (EFT) in \Eq{eq:Shair} is bounded from above as $\Lambda \lesssim 1/|\hat{\alpha}|^{1/2}$, where $\hat{\alpha} = \alpha + i \tilde{\alpha}$.
Since the effects beyond GR (BGR), associated with the scalar hair of black holes, are observable when $|\hat{\alpha}|/r_s^2 = O(1)$, where $r_s$ is the Schwarzschild radius, we find that for phenomenological applications, i.e.~for black holes of astrophysical size, $\Lambda \lesssim 1/\km$.
Therefore, the observability of black holes with scalar hair comes at the high price of a very limited regime of validity of these models.
In fact, we will argue that the observation of $O(|\hat{\alpha}|/\km^2)$ non-standard effects due to the scalar hair of astrophysical black holes is likely at odds with standard gravity at distances shorter than $|\hat{\alpha}|^{1/2}$, or, from a more dramatic perspective, it would point to the violation of fundamental principles below that scale.

Our causality bound is a generalization of the well-known fact that effective field theories exhibiting non-minimal 3-graviton or 2-photon plus 1-graviton interactions, if extrapolated beyond their regime of validity, display time advances when in a gravitational background, in conflict with causality \cite{Drummond:1979pp,Camanho:2014apa}.
In \Sec{sec:timeadv} we show that the scalar-graviton mixing induced by the non-minimal couplings in \Eq{eq:Shair} leads as well to a macroscopic violation of causality unless $\Lambda \lesssim 1/|\hat{\alpha}|^{1/2}$, in which case the time advance is never observable within the EFT regime of validity.
Based on this bound as well as those found in \cite{Camanho:2014apa}, along with the theoretical constraints on gravitational EFTs recently derived using dispersion relations \cite{Bern:2021ppb,Caron-Huot:2022ugt}, we will extract generic lessons on the power counting of gravitational EFT operators, of relevance for gravitational wave science.

While our bound renders the EFT in \Eq{eq:Shair} at the verge of its regime of validity for the physical systems of interest, there is a small range of scales where it could remain interesting. What are the effects one can expect from such a low cutoff? 
In \Sec{sec:beyond} we investigate this question by means of dispersion relations, which connect observables at low energies, i.e.~EFT coefficients, with the high-energy dynamics that lies behind them, on the basis of the unitarity, locality and causality of the scattering amplitudes.
Due to the weakness of the non-minimal gravitational interactions compared to GR, as enforced by causality, we find that our one-loop positivity conditions are not powerful enough to extract a robust answer.
Nevertheless, given that setting $|\hat{\alpha}| \sim 1/\Lambda^2$, so as to maximize the BGR effects within the EFT regime, fixes the power counting of the EFT, we are able to identify in \Sec{sec:ndaexpect} the leading higher-dimensional operators that should generically (yet not generally) become large.
The reader not interested in the more formal discussion of gravitational positivity bounds is invited to directly go to this latter subsection, which is the starting point of our phenomenological analysis.

In \Sec{sec:pheno} we explore the phenomenological consequences of the additional EFT operators. The main generic lesson we extract is that it would be of great significance to extend the black hole solutions and numerical studies of their merger, obtained so far in the literature for the scalar-GB and dynamical-CS gravity theories ($\tilde{\alpha} = 0$ and $\alpha = 0$ respectively), to include these operators.
This conclusion holds insofar there exist a UV completion in which gravity remains well described by GR at scales lower than $|\hat{\alpha}|^{1/2} \sim \km$, an important caveat that we chose to be agnostic about and leave for future investigation. We present our outlook and conclusions in \Sec{conclusions}.

In \App{sec:scalarization} we discuss how our arguments could be extended to place theoretical constraints on the idea of spontaneous scalarization around black holes \cite{Silva:2017uqg,Macedo:2019sem}.

\section{Time advance bounds}
\label{sec:timeadv}

In this section we compute the time delay that the two graviton polarizations and the massless scalar experience when scattering against a very heavy (classical) gravitational source in the eikonal regime, following \cite{Camanho:2014apa,AccettulliHuber:2020oou} to include the effects of the non-minimal couplings in \Eq{eq:Shair}. These interactions lead to a non-diagonal transition amplitude between graviton and scalar, such that one of the propagation eigenmodes travels faster than what is allowed by the causal structure of the asymptotic spacetime, thus violating asymptotic causality \cite{Gao:2000ga}. 
This is analogous to the case of gravitons and photons discussed in \cite{Camanho:2014apa,AccettulliHuber:2020oou}, where non-minimal gravitational 3-point interactions, encoded by the operators $R_{\mu \nu \rho \sigma} R^{\rho \sigma}_{\;\;\;\; \alpha \beta} R^{\alpha \beta \mu \nu}$ and $F_{\mu \nu} F_{\rho \sigma} R^{\mu\nu\rho\sigma}$, give rise to a mixing between the two graviton or the two photon helicities, respectively, and which results in a net macroscopic time advance for one of the propagating eigenmodes.%
\footnote{Similar ideas have been considered for quadratic gravity in \cite{Edelstein:2021jyu}.}
Since this happens for scattering at sufficiently small impact parameters, avoiding causality violation sets an upper bound on the cutoff of the EFT, $\Lambda$, where dynamics that is not captured by the EFT must become relevant.
For recent works discussing the notion of causality in the gravitational context, we point the reader to e.g.~\cite{Goon:2016une,deRham:2019ctd,deRham:2020zyh,Bellazzini:2021shn,Chen:2021bvg,deRham:2021bll}.

Let us then consider the scattering of graviton and scalar with an spectator of mass $m$, within the so-called eikonal limit, $s \gg t$, where $s$ is the center of mass energy of the collision and $t = - |\vec{q}|^2 \equiv q^2$, where $\vec{q}$ is the exchanged momentum. 
We take the spectator to be very heavy and nearly at rest, acting as a gravitational source against which the massless probe particle, of energy $\omega$, scatters. 
In such a kinematical configuration, $m \gg \omega \gg q$, the leading contribution to the gravitational amplitude for $r_s \omega > 1$, with $r_s = m/(4\pi\mpl^2)$ the Schwarzschild radius of the target, is obtained after summing over ladder diagrams from single graviton exchange, see \Fig{fig:4pt}. The $S$-matrix takes an exponential form, $S = e^{i \delta(\omega,b)}$, where 
\beq
\label{eq:delta}
	\delta(\omega,\vec{b}) = \dfrac{1}{4m\omega}\int\dfrac{d^{D-2}q}{(2\pi)^{D-2}}e^{i\vec{q}\cdot\vec{b}}\amp(\omega,\vec{q})\,,
\eeq
is the eikonal phase shift and $\vec{b}$ the impact parameter \cite{Amati:1992zb,Kabat:1992tb}.
As we show below, the phase shift is in general a matrix in helicity space, from which, after diagonalization, one can extract the classical time delay for the propagation eigenmodes simply as $\Delta t=\partial_{\omega}\delta$.%
\footnote{One could consider as well, as done in \cite{Camanho:2014apa}, the sub-planckian scattering against a coherent state of a large number $N \gg 4 \pi \mpl^2/s$ of relativistic particles, a.k.a.~shock waves.}

Let us briefly go over the time delay for a probe particle minimally coupled to gravity, that is the Shapiro time delay.
The tree-level amplitude is helicity-preserving and universal,
\begin{align}
\label{eq:ampGR}
	\amp^{\GR}_{\textrm{tree}} \simeq \dfrac{1}{\mpl^2}\dfrac{(2m\omega)^2}{q^2}\,.
\end{align}
We can compute the associated phase shift by performing the integral \Eq{eq:delta} in $D-2$ dimensions, where $D=4-2\epsilon$ is used as a regularization,
\begin{align}
\label{eq:deltaGR}
	\delta^{\text{GR}} = \dfrac{m\omega}{4\pi\mpl^2}\Gamma\left(\dfrac{D-4}{2}\right)\dfrac{1}{b^{(D-4)/2}}=2\omega r_s\left(-\dfrac{1}{2\epsilon}-\dfrac{\gamma_E}{2}-\log b\right)+O(\epsilon)\,,
\end{align}
where $b \equiv |\vec{b}|$.
Subtracting the time delay measured at a reference impact parameter $b_0 \gg b$, we obtain the result, 
\begin{align}
\label{eq:shapiro}
	\Delta t_{\GR}=2r_s\log(b_0/b)\,.
\end{align}
This is the Shapiro time delay for a signal travelling at an impact parameter $ b $ from a source with Schwarzschild radius $r_s$, as measured by an observer at an impact parameter $ b_0\gg b $.

Within GR, the leading corrections to the phase shift are of order $r_s/b$, 
associated to amplitude terms in momentum space of order $q/\omega$, arising from the eikonal expansion as well as non-linear gravitational interactions \cite{Akhoury:2013yua}. Note that when these corrections become large, that is when $r_s \sim b$, the deflection angle of the probe, $\theta = - \omega^{-1}  \partial_{b}\delta$, is no longer small.
Let us point out as well that as long as $r_s \omega > 1$, the Shapiro time delay is larger than the quantum-mechanical uncertainty associated with the probe wave, i.e.~$\Delta t_{\GR} > 1/\omega$.

\subsection{Non-minimal scalar-tensor trilinear interactions}
\label{sec:scalartensor}

The (pseudo)scalar-graviton 3-point interactions associated with the $\phi \mathcal{R}_{\GB}^2$ and $\phi R \tilde{R}$ operators in \Eq{eq:Shair} give rise to an eikonal phase shift that is not diagonal with respect to the helicity of the probe particle. This, along with the energy dependence of the interaction, results in time advances at energies where the EFT is still weakly coupled.

\begin{figure}[tb]
	\centering
	\includegraphics[width=0.65 \textwidth]{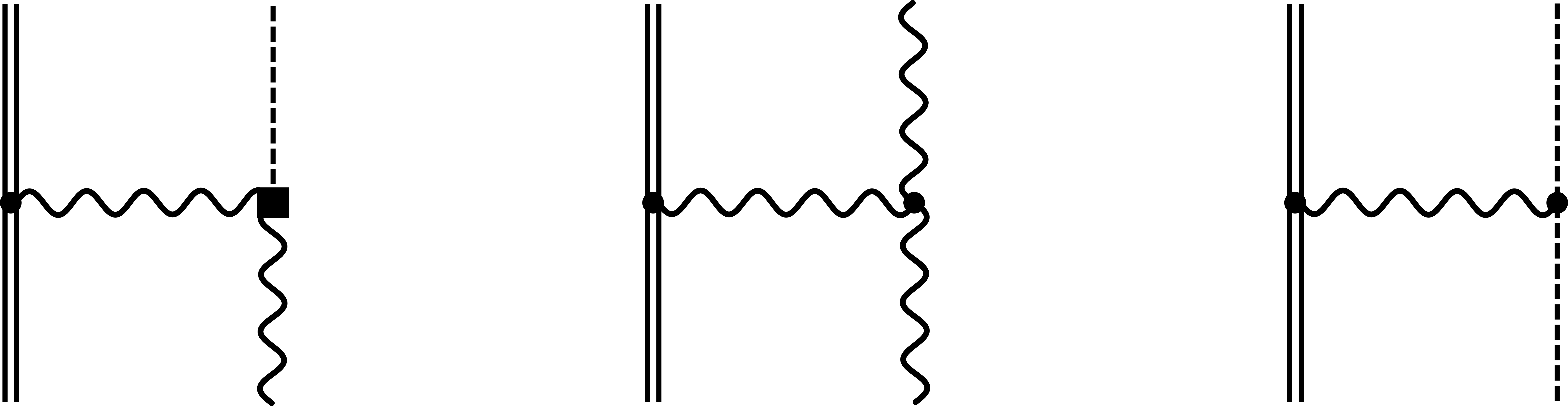}
	\caption{\footnotesize Leading tree-level diagrams for the eikonal scattering of graviton and scalar against a heavy target. Wiggly lines represent gravitons, dashed lines the massless scalar, 
	and double lines the massive source. The square vertex corresponds to the $\phi RR$ helicity-changing interaction.
	}
	\label{fig:4pt}
\end{figure}

In order to compute the phase shift, we consider 4-point scattering amplitudes associated with tree-level graviton exchange between a scalar or graviton and a heavy spectator, which we take to be a scalar,  $S$, without loss of generality. The corresponding Feynman diagrams are shown in \Fig{fig:4pt}.

Using spinor-helicity variables and taking all the particles (with complex momenta) as incoming, the relevant 3-point amplitudes read as follows:
\beq
\label{eq:3pointGB}
\amp^{\GB/\CS}_{1_\phi 2_{h^{++}} 3_{h^{++}}} = \frac{2 \hat{\alpha}}{\mpl} [23]^4 \, , \quad \amp^{\GB/\CS}_{1_\phi 2_{h^{--}} 3_{h^{--}}} = \frac{2 \hat{\alpha}^*}{\mpl} \langle 23 \rangle^4 \, ,
\eeq
where recall that $\hat{\alpha} = \alpha +i \tilde{\alpha}$ and that $\hat{\alpha} = 0$ and $\alpha = 0$ correspond  to the scalar Gauss-Bonnet and dynamical Chern-Simons gravity theories, respectively.
In the regime $m \gg \omega \gg q$, the relevant $4$-point scattering amplitudes are given by
\begin{align}
	\amp^{\GB/\CS}_{1_S 2_S 3_{h^{++}} 4_{\phi}}&=\amp^{\GB/\CS}_{1_S 2_S 3_{\phi} 4_{h^{++}}}\simeq -\frac{2\hat{\alpha}}{\mpl^2} \dfrac{(q_1+iq_2)^2}{q^2}(2m\omega)^2\,,\\
	\nonumber \amp^{\GB/\CS}_{1_S 2_S 3_{h^{--}} 4_{\phi}}&=\amp^{\GB/\CS}_{1_S 2_S 3_{\phi} 4_{h^{--}}}\simeq -\frac{2\hat{\alpha}^*}{\mpl^2} \dfrac{(q_1-iq_2)^2}{q^2}(2m\omega)^2\,,
\end{align}
where $ q_1,q_2 $ are the components of the exchanged momentum $\vec{q}$.
Defining $ b_{\pm}=({b}_1\pm i{b}_2)/2$, we have $ \vec{b}\cdot\vec{q}=b_{+}(q_1-iq_2)+b_{-}(q_1+iq_2) $, and as before the eikonal phase shift matrix is obtained by taking the impact-parameter transform of the amplitudes,
\begin{align}
	\delta^{\GB/\CS}_{1_S 2_S 3_{h^{++}}4_{\phi}}=\delta^{\GB/\CS}_{1_S 2_S 3_{\phi} 4_{h^{++}}}=-2\omega r_s\,\dfrac{\hat{\alpha}}{b_-^2} \,,\\
	\nonumber \delta^{\GB/\CS}_{1_S 2_S 3_{h^{--}}4_{\phi}}=\delta^{\GB/\CS}_{1_S 2_S 3_{\phi} 4_{h^{--}}}=-2\omega r_s\,\dfrac{\hat{\alpha}^*}{b_+^2} \,.
\end{align}
These helicity-changing contributions add up to the helicity-preserving ones from minimal coupling, to yield the phase shift matrix
\begin{align}
\label{eq:elementsdelta}
	\delta^{\GR+\GB/\CS}\simeq2\omega r_s\begin{pmatrix}
		D&0&A\\
		0&D&A^*\\
		A^*&A&D\\
	\end{pmatrix},
\end{align}
with rows $(h^{++}, \, h^{--}\, , \phi)$ and 
\begin{align}
	D=-\dfrac{1}{2\epsilon}-\dfrac{\gamma_E}{2}-\log b\,, \quad  
	A=-\dfrac{\hat{\alpha}}{b_-^2}\,.
\end{align}
After diagonalizing, we find the eigenvalues
\beq
	\delta_0=2\omega r_s\left(-\dfrac{1}{2\epsilon}-\dfrac{\gamma_E}{2}-\log b\right)\,, \quad \delta_{\pm}=2\omega r_s\left(-\dfrac{1}{2\epsilon}-\dfrac{\gamma_E}{2}-\log b\pm\sqrt{2}\dfrac{|\hat{\alpha}|}{b^2}\right) \, ,
\eeq
where the first corresponds to a pure graviton state, while the other two to a scalar-graviton mixed state. 
The time delay that the latter propagating eigenmodes acquire are
\beq
\label{eq:timedel}
	\Delta t_{\pm}=2r_s\left(\log \dfrac{b_0}{b}\pm\sqrt{2}\dfrac{|\hat{\alpha}|}{b^2}\right)\,.
\eeq
At small enough impact parameters, $\Delta t_{-}$ becomes negative, that is a time advance, signalling a potential violation of causality. Phrased in another way, for a given impact parameter there is a time advance if the GB/CS coefficient is large enough, $|\hat{\alpha}| \gtrsim b^2 \log(b_0/b)$.
To avoid acausality at low energies, the EFT computation must therefore break down at distances such that this condition cannot be satisfied.%
\footnote{That is, at a distance $1/\Lambda > b_*$, where the lasrgest impact parameter at which a time advance is found, $b_*$, is given by $|\hat{\alpha}| \sim b_*^2 \log(b_0/b_*)$. Note that the dynamics needed to restore causality on times cales of order $b_*$ should only involve momentum transfers $q_* \sim 1/b_*$, meaning that new physics must appear at scales $1/b_*$ or smaller, a priori regardless of $\omega$ and $r_s$.}
This implies the GB/CS coupling is parametrically bounded by the cutoff of the EFT as
\beq
\label{eq:boundGB}
	|\hat{\alpha}| \lesssim \dfrac{\log(b_0 \Lambda)}{\Lambda^2} \,.
\eeq

Several comments are in order. For the violation of causality to potentially be resolvable and thus become problematic, the time advance should be larger than the quantum uncertainty of the wave-packet, $|\Delta t_{-}| > 1/\omega$. 
For impact parameters where the BGR contribution is assumed to dominate, this condition reads%
\footnote{This condition can also be interpreted as the one for which the  beyond-GR contribution to the time delay is resolvable on its own.}
\beq
\label{eq:resolvable}
	|\Delta t_{-}| \sim r_s \dfrac{|\hat{\alpha}|}{b^2} > \dfrac{1}{\omega} \, ,
\eeq
which for impact parameters down to the minimum cutoff length implied by \Eq{eq:boundGB}, i.e.~$b \sim |\hat{\alpha}|^{1/2}$ (neglecting the $\log$), just requires $r_s \omega > 1$.
Equivalently, \Eq{eq:resolvable} defines an impact parameter below which the would-be time advance is resolvable, $b_r = (r_s \omega |\hat{\alpha}|)^{1/2}$. 
This is larger than the minimum cutoff length within the EFT regime of validity, i.e.~$b_r > |\hat{\alpha}|^{1/2}$, as long as $r_s \omega > 1$. Therefore, even if potentially resolvable, as long as \Eq{eq:boundGB} is satisfied there is never an actual time advance.
Alternatively, one could also argue that the time advance is actually not resolvable at $b \sim |\hat{\alpha}|^{1/2}$ because the UV completion precludes $r_s \omega > 1$, which in practice means a cutoff such that $\omega \lesssim \Lambda \lesssim 1/r_s$. In this case, the condition $|\Delta t_{-}| < 1/\omega$ for $\omega \sim \Lambda$ and $b \sim r_s \sim 1/\Lambda$ implies $\Lambda \lesssim |\hat{\alpha}|^{1/2}$, just like in \Eq{eq:boundGB} up to $O(1)$ factors, which in any case we are oblivious about.

The bound \Eq{eq:boundGB} depends on the logarithm of an unspecified scale $b_0 \gg b$, because of the infrared (IR) divergent nature of gravity in four dimensions.
While the identification of IR-finite scattering observables in gravity remains an open and interesting problem (see e.g.~\cite{Bellazzini:2019xts} for a recent attempt in the context of causality constraints), we do not regard this divergence as a serious drawback that invalidates our bounds. In fact, note that even if we consider an IR scale of the order of the size of the observable universe, we merely find $\log(\Lambda/H_0) \sim 50$.

Finally, let us point out that $\Lambda \lesssim |\hat{\alpha}|^{-1/2}$ implies that the EFT description must break down at energies much lower than the strong coupling scale associated with the GB/CS interactions. Indeed, the scale where the trilinear scalar-tensor coupling becomes large, indicated by e.g.~the 4-graviton amplitude mediated by the scalar becoming strong, $\amp \sim (\hat{\alpha}/\mpl)^2 E^6 \sim 1$, is
\beq
\label{eq:Lalpha} 
\La = \left(\dfrac{\mpl}{|\hat{\alpha}|}\right)^{1/3} \, ,
\eeq
much larger than the actual cutoff of the EFT, unless $|\hat{\alpha}| \sim 1/\mpl^2$.

\subsection{Causality bounds on power counting}
\label{sec:ndabounds}

In this section we reinterpret the causality constraints in terms of bounds on the power counting of gravitational EFTs.
With this aim, let us consider the generic form of a scalar theory coupled to gravity, in which the heavy degrees of freedom, of mass $\Lambda$ or higher (i.e.~$\Lambda$ is the EFT cutoff), have been integrated out
\beq
\Lag = \tfrac{1}{2} \hat{M}_{\rm Pl}^2 R + \frac{\Lambda^4}{g^2} 
L^{(0)} \left( \frac{\nabla_\mu}{\Lambda}, \frac{\zeta R_{\mu \nu \rho \sigma}}{\Lambda^2}, \frac{g\phi}{\Lambda} \right) + \dots \, .
\label{eq:NDA}
\eeq
In the spirit of naive dimensional analysis (NDA) each covariant derivative $\nabla_\mu$ is weighted by $1/\Lambda$, and each (scalar) field $\phi$ by $g/\Lambda$.
The coupling $g$ parametrizes the strength with which the heavy states couple to the light degrees of freedom, with $g \sim 4 \pi$ the usual non-perturbative coupling limit.
Note that instead of considering the Riemann tensor, $R_{\mu \nu \rho \sigma} \sim \partial_\mu \partial_\nu h_{\rho \sigma}$, simply as a two-derivative object thus weighted by $1/\Lambda^2$, we introduce a dimensionless parameter $\zeta$ to allow for the possibility that gravitational interactions beyond GR's minimal coupling are enhanced w.r.t.~standard NDA.%
\footnote{As usual we work with a dimensionless graviton field, whose interactions are eventually weighted by $1/\mpl$ once its kinetic term is canonically normalized, following the normalization of the Einstein-Hilbert (EH) term.}
We will elaborate on such a generalized power counting below.
Each $\phi$ interaction comes with a decay constant $f$, identified with (or defined as)
\beq
f = \dfrac{\Lambda}{g} \,.
\label{eq:decayconstant}
\eeq

At this point we can already distinguish the two interesting scenarios, for which it is enough to consider the standard power counting $\zeta = 1$ and to realize that the EH action receives a contribution from both terms in \Eq{eq:NDA}. When $\hat{M}_{\rm Pl}^2 \gg f^2$, the EH action is dominated by the first term and the effective Planck scale is $\mpl \sim \hat{M}_{\rm Pl}$. Gravity is external to the ultraviolet (UV) dynamics giving rise to $L^{(0)}$, a.k.a.~``elementary''. Instead, when $\hat{M}_{\rm Pl}^2 \ll f^2$, we have $\mpl \sim f$ and the heavy dynamics constitutes a bona fide UV completion of gravity. Phrasing it in terms of the coupling $g$, the  minimum coupling $g \sim \Lambda/\mpl$ corresponds to the ``composite'' limit of gravity. 
This is the case of string theory (or more generally, potential tree-level UV completions with infinitely many higher-spins particles, see e.g.~\cite{Camanho:2014apa, Kologlu:2019bco}), where $\Lambda \sim M_s$ the string scale, as well as of loop-level completions based on a large number of species, $N \sim (4 \pi \mpl/\Lambda)^2$, where $g \sim 4 \pi/\sqrt{N}$ \cite{Veneziano:2001ah,Dvali:2007hz}.
Note that in this limit one finds the largest coefficients for gravitational EFT operators with none or a single matter field, since they scale as $1/g^2$ or $1/g$ respectively. The fact that $g \gtrsim \Lambda/\mpl$ is reminiscent of the weak gravity conjecture \cite{Arkani-Hamed:2006emk}.

In terms of scattering amplitudes, the two scenarios are distinguished by the maximal size of e.g.~2-to-2 graviton processes within the EFT regime of validity, i.e.~$E \lesssim \Lambda$. From minimal coupling we have $\amp^\GR \sim (E/\mpl)^2 \lesssim (\Lambda/\mpl)^2$. Instead, an effective operator like $R_{\mu \nu \rho \sigma}^{\, 3}$ leads to an amplitude $\amp^\BGR_{(\zeta = 1)} \sim E^6/(g^2 \Lambda^2 \mpl^4) \lesssim f^2 \Lambda^2/\mpl^4$, smaller than GR except in the limit $\mpl \sim f$, in which case the two amplitudes are of the same size at the cutoff.
The same analysis can be reproduced if instead of amplitudes one considers other (classical) gravitational observables and distances, rather than energies, within EFT control, i.e.~$r \gtrsim 1/\Lambda$.

For the generalized power counting $\zeta > 1$, the discussion is very much analogous, except for the important difference that now the BGR effects can become larger than the GR prediction for energies well described by the EFT.
The composite case corresponds to $\hat{M}_{\rm Pl}^2 \ll \zeta f^2$, for which we have $\mpl \sim \sqrt{\zeta} f$. Therefore, $\zeta \ll (\mpl/f)^2$ corresponds to the case where gravity is external to the UV dynamics. Elementary or composite, we find that non-standard gravitational interactions, in the form of $R_{\mu \nu \rho \sigma}^3$, give rise to enhanced 4-graviton amplitudes
\beq
\amp^\BGR_{1_h 2_h 3_h 4_h} \sim \zeta^3 \dfrac{E^6 f^2}{\Lambda^4 \mpl^4} \lesssim \zeta^3 \dfrac{\Lambda^2 f^2}{\mpl^4} \lesssim \dfrac{\Lambda^2 \mpl^2}{f^4} = \left( \dfrac{g \mpl}{f} \right)^2\, ,
\label{eq:maxamp}
\eeq
where the first inequality follows from $E \lesssim \Lambda$ and the second from $\zeta \lesssim (\mpl/f)^2$. Note that for $\zeta \gtrsim (\mpl/f)^{2/3}$ the amplitude is larger than in GR, and it becomes non-perturbatively strong, i.e.~$\amp \sim (4 \pi)^2$, for EFT cutoffs well below the maximal gravity cutoff given by $4\pi \mpl$. As we discuss in the following, it is precisely this possibility that causality constraints forbid.%
\footnote{It is perhaps instructive to compare with EFTs for spin-1 (abelian or non-abelian) gauge fields,
\beq
\Lag = \tfrac{1}{4 \hat{e}^2} F_{\mu \nu}^2 + \frac{\Lambda^4}{g^2} 
L^{(0)} \left( \frac{D_\mu}{\Lambda}, \frac{\zeta F_{\mu \nu}}{\Lambda^2}, \frac{g\phi}{\Lambda} \right) \, ,
\nonumber
\eeq
with the elementary and composite limits given respectively by $\hat{e} \ll g/\zeta$ (and effective gauge coupling $e \sim \hat{e}$) and $\hat{e} \gg g/\zeta$ ($e \sim g/\zeta$). As discussed in \cite{Liu:2016idz}, in the strongly coupled gauge field scenario $\zeta \sim g/e \gtrsim 1$ one finds 4-point amplitudes (from e.g.~$F_{\mu \nu}^4$ operators) $\amp \sim g^2 (E/\Lambda)^4$, which can be larger than the amplitude from minimal coupling, $\amp \sim e^2$, for energies within the EFT.
}

Let us start by recalling that each specific UV theory within the class of theories described in the IR by \Eq{eq:NDA} comes with $O(1)$ factors not captured by the power counting.
Even more importantly, the presence of symmetries can enforce some EFT operators to have vanishing coefficients, for instance if $\phi$ is Nambu-Goldstone boson with a shift symmetry $\phi \to \phi + c$ (as the scalar field that concerns us in this work), any potential term for $\phi$ vanishes.
However, beyond the well-known selection rules from symmetries, there are further requirements that an EFT must satisfy if it is to be consistent with the fundamental principles of unitarity, locality, and causality (and if it is to arise from UV dynamics that abides by such principles).
Indeed, it was found in \cite{Camanho:2014apa} that causality, in the form of absence of a (resolvable) time advance, leads to a constraint on the size of corrections to the cubic graviton coupling, arising from an operator $\alpha_3 \mpl^2 R_{\mu \nu \rho \sigma}^3$, given by $\alpha_3 \lesssim 1/\Lambda^4$. In terms of the power counting \Eq{eq:NDA}, $\alpha_3 \sim \zeta^3/(g \mpl \Lambda)^2$, such a bound implies $\zeta \lesssim (\mpl/f)^{2/3}$, precisely such that the BGR effects never get to dominate over GR, see below \Eq{eq:maxamp}.
This conclusion seems to be generic. 
The similar bound we have derived in \Sec{sec:scalartensor} on the GB/CS non-minimal coupling of gravitons to a scalar, $|\hat{\alpha}| \lesssim 1/\Lambda^2$, when interpreted in terms of our power counting, $|\hat{\alpha}| \sim \zeta^2/(g \mpl \Lambda)$, implies $\zeta \lesssim (\mpl/f)^{1/2}$.
Once again, this forbids the 4-graviton amplitude mediated by the scalar from getting larger than in GR if restricted to energies within the EFT, $E \lesssim \Lambda$,
\beq
\amp^{\GB/\CS}_{1_{h^{++}} 2_{h^{++}} 3_{h^{--}} 4_{h^{--}}} \sim |\hat{\alpha}|^2 \dfrac{E^6}{\mpl^2} 
\lesssim \dfrac{\Lambda^2}{\mpl^2} \, .
\label{eq:maxampGB}
\eeq
It is illuminating to realize that in the case of a standard power counting $\zeta = 1$, these causality constraints robustly imply that $g \gtrsim \Lambda/\mpl$ (or equivalently $f \lesssim \mpl$), as we expected from the simple NDA considerations on the elementary vs composite nature of gravity. 
In turn, if one is interested in genuine UV completions of gravity, i.e.~$g \sim \Lambda/\mpl$, these bounds imply that $\zeta \lesssim 1$ and therefore that the EFTs in which non-minimal interactions are enhanced beyond standard NDA have no gravitational completions consistent with fundamental principles.

\subsubsection{Bounds from dispersion relations}

This conclusion is reinforced by recent progress on the derivation of theoretical constraints on gravitational EFTs that go beyond causality violation in classical observables and therefore beyond corrections to cubic gravitational interactions \cite{Bellazzini:2015cra,Hamada:2018dde,Bellazzini:2019xts,Chowdhury:2019kaq,Tokuda:2020mlf,Alberte:2020bdz,Arkani-Hamed:2020blm,Caron-Huot:2021rmr,Bern:2021ppb,Arkani-Hamed:2021ajd,Chowdhury:2021ynh,Caron-Huot:2022ugt,Chiang:2022jep}. Such bounds are instead obtained via dispersion relations \cite{Gell-Mann:1954ttj}, which connect the coefficients of the EFT operators to the dynamics of their UV completions. These UV/IR relations, which we will review in some detail in \Sec{sec:beyond}, are very powerful because of their generality, relying only on the basic assumptions of unitarity, locality and causality (encoded as the analyticity, crossing symmetry and boundedness of the scattering amplitudes).%
\footnote{The link between dispersion relations and causality, expressed as the absence of superluminal propagation, was pointed out in \cite{Adams:2006sv}, and its connection with time delay has been recently discussed in \cite{Arkani-Hamed:2020blm,Caron-Huot:2022ugt}.}
Of particular relevance for the physics of black holes are the results of \cite{Bern:2021ppb}, which derived a lower bound on $\alpha_4 \mpl^2 R_{\mu \nu \rho \sigma}^4$ given by $\alpha_4 \gtrsim \alpha_3^2 \Lambda^2$ (recall $\alpha_3 \mpl^2 R_{\mu \nu \rho \sigma}^3$), and of \cite{Caron-Huot:2022ugt}, which derived the upper bound $\alpha_4 \lesssim 1/\Lambda^6$. Both constraints restrict the BGR contribution to gravitational observables to be smaller than the prediction of GR. 
In fact, we should stress that if similar bounds were to be derived on non-standard higher-point amplitudes (with $n \geqslant 5$ gravitons) from $R_{\mu \nu \rho \sigma}^n$ operators, we would be led to the conclusion that the power counting in \Eq{eq:NDA} with $\zeta > 1$ is inconsistent altogether, i.e.~regardless of $f$ and not only for $f \sim \mpl$.
While this seems like a plausible expectation, a robust derivation of theoretical constraints on higher-point amplitudes remains an open problem at the time of writing this work (see e.g.~\cite{Chandrasekaran:2018qmx}  for recent progress in this direction).
If indeed $\zeta > 1$ is forbidden by fundamental principles, we would come to the sensible conclusion that in a gravitational EFT the largest effects for a fixed cutoff are found when $f \sim \mpl$, therefore when gravitational interactions should dramatically change above $\Lambda$. We will provide further insight into this fact in \Sec{conclusions}.

Before concluding this section, let us make some additional comments on the implications of the causality bounds on the phenomenology of black holes beyond GR. 
The constraint $\alpha_4 \lesssim 1/\Lambda^6$ \cite{Caron-Huot:2022ugt} places modifications of GR due to quartic terms in the curvature on the same footing as those due to cubic terms.
This means that there is no strong reason to discard the effects of $R_{\mu \nu \rho \sigma}^3$ operators, leading in the derivative expansion, while keeping those of $R_{\mu \nu \rho \sigma}^4$ \cite{Endlich:2017tqa,Cardoso:2018ptl,Sennett:2019bpc}.

The constraint we have obtained in \Sec{sec:scalartensor} on BGR scalar-tensor cubic couplings have also been recently derived in \cite{Caron-Huot:2022ugt} via dispersion relations. In this regard, it is important to point out that even though our bound is robust up to $O(1)$ factors, contrary to the more precise (yet still IR divergent) one from dispersion relations, we believe our derivation is very valuable because it comes from a simple physical setup in which causality violation is a classical, macroscopic effect, therefore it does not rely on a priori stronger assumptions on the analyticity and polynomial boundedness of scattering amplitudes associated with causality.

NDA expectations are also confirmed by dispersion relations involving operators with extra derivatives acting on the curvature, for instance of the form 
\beq
\alpha_5 \mpl^2 R_{\mu \nu \rho \sigma}^2 (\nabla_\eta R_{\mu \nu \rho \sigma})^2,\quad  \alpha_6 \mpl^2 (\nabla_\eta R_{\mu \nu \rho \sigma})^4,
\eeq
which contribute to 4-graviton amplitudes as $\amp \sim \alpha_J E^{2J}$ \cite{Bern:2021ppb,Caron-Huot:2022ugt,Chiang:2022jep}. In particular, there are an infinite number of linear constraints on the EFT coefficients that take the form of two-sided bounds such as 
\beq
-\alpha_4 \leqslant \alpha_5 \Lambda^2 \leqslant \alpha_4 \quad \mathrm{and} \quad 0 \leqslant \alpha_6 \Lambda^4 \leqslant \alpha_4\;,
\eeq
and similar bounds for higher $ J $, respectively odd or even.

Their interpretation in terms of a power counting is clear, subleading operators in the derivative expansion $\nabla/\Lambda$ cannot be enhanced over the leading ones.

In our discussion we have focussed on cubic and quartic operators built out of the Riemann tensor, with no mention of terms quadratic in the curvature. 
This is because $R_{\mu \nu \rho \sigma}^{\, 2}$ operators do not contribute to graviton scattering amplitudes, given that the GB term ${\cal R}^2_{\rm GB}$ is a topological invariant (in $D = 4$) and because field redefinitions can be performed to eliminate any EFT operator built out of $R$ and $R_{\mu \nu}$ in favour of matter terms ($T$ and $T_{\mu \nu}$), therefore giving rise to amplitudes involving $\phi$ fields. For this reason, one might find it more convenient (although not necessary) to use a basis of EFT operators directly link to scattering amplitudes, such as the one systematically constructed in \cite{Ruhdorfer:2019qmk}.
In this respect, note that the relevant object giving rise to processes with gravitons on-shell is the Weyl tensor, $C_{\mu \nu \rho \sigma} = R_{\mu \nu \rho \sigma} - (g_{\mu [ \rho} R_{\sigma]\nu}  -g_{\nu [ \rho} R_{\sigma]\mu}) + \tfrac{1}{3} g_{\mu [\rho} g_{\sigma]\nu} R$. This means that the relevant parts of the GB/CS operators in \Eq{eq:Shair} are 
$\mpl \alpha \, \phi C_{\mu\nu\rho\sigma} C^{\mu\nu\rho\sigma}$ 
and $\mpl \tilde{\alpha} \, \phi C_{\mu\nu\rho\sigma} \tilde C^{\mu\nu\rho\sigma}$. 
These are also the terms behind the scalar hair of black holes, since black holes are Ricci-flat gravitational solutions ($R, R_{\mu \nu} = 0$) at zeroth order in $\alpha$, $\tilde{\alpha}$.

Finally, causality bounds on pure scalar operators are also relevant for the physics of hairy black holes, in particular 
\beq
\tfrac{1}{4} c_2 (\nabla_\mu \phi)^4\;,
\eeq 
i.e. the leading operator in the derivative expansion.
Several recent works on dispersion relations that incorporate gravity have argued that $c_2 \gtrsim - 1/(\Lambda^2 \mpl^2)$ (and likewise for the equivalent operator $F_{\mu\nu}^4$ in a theory of photons) \cite{Hamada:2018dde,Bellazzini:2019xts,Tokuda:2020mlf,Alberte:2020bdz,Caron-Huot:2021rmr,Henriksson:2022oeu}, bound that becomes a standard positivity constraint, $c_2 > 0$ \cite{Adams:2006sv}, when gravity decouples, $\mpl \to \infty$.
In particular, \cite{Caron-Huot:2021rmr} has shown via dispersion relations at finite impact parameter that this is indeed the case up to a $\log(b_0 \Lambda)$, as in \Eq{eq:boundGB}.
Furthermore, the upper bound $c_2 \lesssim (4 \pi)^2/\Lambda^4$ has been derived using similar techniques \cite{Caron-Huot:2020cmc,Du:2021byy}.
These constraints on $c_2$ can be easily understood in terms of the power counting in \Eq{eq:NDA}. Since $c_2 \sim g^2/\Lambda^4$, the upper and lower bounds correspond, respectively, to the maximum coupling in the spirit of NDA, $g \lesssim 4\pi$, and to the minimum coupling in a gravitational theory, $g \gtrsim \Lambda/\mpl$ (recall that power counting estimates are insensitive to the sign of the operators' coefficient).


\section{Signs of UV completion}
\label{sec:UV}

In the previous section we have argued that gravitational EFTs where black holes have scalar hair, \Eq{eq:Shair}, must have a cutoff $\Lambda \lesssim |\hat{\alpha}|^{-1/2}$. In terms of the power counting \Eq{eq:NDA} with $\zeta = 1$, the maximum cutoff of these theories corresponds to the minimum NDA coupling $g \sim \Lambda/\mpl$, while EFTs with a larger coupling, or equivalently $f < \mpl$, must have a lower cutoff for the same value of $|\hat{\alpha}|$.

In this section we try to infer from an EFT point of view what additional low-energy effects are associated with a generic UV completion at the scale $\Lambda$, in particular one that is unitary, local and casual.
We focus on the leading corrections in the derivative and field expansion \cite{Ruhdorfer:2019qmk}, restricted to CP even operators
\beq
\Delta S = \int d^4 x \sqrt{-g} \left[ \mpl^2 \left( \alpha_3 \mathcal{I} + \alpha_4 \mathcal{C}^2 + \alpha'_4 \tilde{\mathcal{C}}^2 \right) + \dfrac{c_2}{4} (\nabla_\mu \phi)^4 + \dfrac{d_1}{2} \mathcal{C} (\nabla_\mu \phi)^2  \right] \, ,
\label{eq:Sadd}
\eeq
where $\mathcal{C} = R_{\mu \nu \rho \sigma} R^{\mu \nu \rho \sigma}$, $\tilde{\mathcal{C}} = R_{\mu \nu \rho \sigma} \tilde{R}^{\mu \nu \rho \sigma}$ and $\mathcal{I} = R_{\mu \nu}^{\quad \rho \sigma} R^{\mu \nu \alpha \beta} R_{\alpha \beta \rho \sigma}$.
Note that the last operator is equivalent, by the leading-order scalar equation of motion, to the cubic Galileon term $(\nabla \phi)^2 \Box \phi$.
We first investigate the constraints on the coefficients above that arise from dispersion relations at one loop, which take the form of lower bounds that depend on $|\hat{\alpha}|$.%
\footnote{There are no constraints of this form from dispersion relations at tree level. This is in contrast with the lower bound $\alpha_4 \gtrsim \alpha_3^2 \Lambda^2$ \cite{Bern:2021ppb}.}
Precisely because of the upper bound $|\hat{\alpha}| \lesssim 1/\Lambda^2$, we find that such dispersion relations are in fact dominated by standard gravitational contributions, rendering the constraints on the operators in \Eq{eq:Sadd} inapplicable and phenomenologically irrelevant. 
We therefore leave aside general bounds and turn to generic expectations based on power counting.
We show how typical UV completions of the scalar-GB or dynamical-CS theories likely give rise to higher-curvature terms of the same parametric size, i.e.~$\alpha \Lambda^2 \sim \alpha_3 \Lambda^4 \sim \alpha_4 \Lambda^6$, if these arise at the same loop order.

\subsection{Beyond positivity constraints}
\label{sec:beyond}

There exists an extensive literature on dispersion relations, in particular on non-gravitational theories, with many new results and applications found in recent years. We refer the reader to e.g.~\cite{Adams:2006sv,Vecchi:2007na,Bellazzini:2014waa,Bellazzini:2016xrt,deRham:2017avq,deRham:2017zjm,Tolley:2020gtv,Arkani-Hamed:2020blm,Bellazzini:2020cot} and references therein for many of the details we will not present here. 
Dispersion relations are typically constructed by evaluating a 2-to-2 scattering amplitude $\amp(s,t)$ over a closed circular contour in the complex $s$-plane,%
\footnote{In a slight abuse of notation, we denote the amplitude as a function of the $s = - (p_1 + p_2)^2$ and $t = - (p_1+ p_3)^2$ Mandelstam variables as $\amp$, like in \Sec{sec:timeadv} where instead was a function of $\omega$ and $\vec{q}$. Besides, we work with all momenta incoming and recall $u = - (p_1 + p_4)^2 = -s-t + 4 m^2$, where $m$ is now the mass of the scattered states, that we will eventually take to zero.}
\beq
\Sigma_n(s,t) = \frac{1}{2 \pi i} \oint_{\Gamma_s} ds' \frac{\amp(s',t)}{\left(s'+\frac{t}{2}\right)^{n+1}} \, .
\label{eq:dispersion}
\eeq
Let us start with the scattering of the GB/CS scalar $\phi$ at low energies, $s \ll \Lambda^2$, neglecting for the time being GR's minimal gravitational coupling. The 4-scalar EFT interaction in \Eq{eq:Sadd} leads to an amplitude,
\beq
\amp^{\Delta S}_{1_\phi 2_\phi 3_\phi 4_\phi} (s,t) = \dfrac{c_2}{2} (s^2 + t^2 + u^2) \, ,
\label{eq:phi4c2}
\eeq
which grows like $s^2$ for fixed $t$. Therefore, considering a small contour $\Gamma_0$ 
around $s' = -t/2$, the integral of the twice-subtracted amplitude, i.e.~$n=2$ in \Eq{eq:dispersion}, yields $\Sigma_2(0,t) = c_2$
At this point, unitarity and causality allow one to deform the contour away from the origin (for $0 \leqslant t \leqslant 4 m^2$) in a controlled way. First, they imply that $\amp(s,t)$ is analytic everywhere except in the real axis, where one finds singularities in the form of simple poles and branch cuts. The former correspond to particles exchanged at tree level going on shell, which in the case at hand belong only to the UV completion, either in the $s$-channel at $s \geqslant \Lambda^2$ or in the $u$-channel $s \leqslant -\Lambda^2 -t$.
The branch cuts are associated with logarithms arising from loops and correspond to multi-particle production. Besides the loops of heavy states at and above the cutoff, there is an $s$-channel branch cut starting at $s  = 4 m^2$ from (one) loop diagrams of the IR degrees of freedom, and its $s \leftrightarrow u$ crossing symmetric counterpart.
Because of real analyticity, $\amp^*(s,t) = \amp(s^*,t)$, these discontinuities are proportional to ${\rm Im}\amp(s,t)$, which is positive for elastic scattering around $t = 0$. In particular, for the zeroth-order term in an expansion around the forward limit, the optical theorem fixes ${\rm Im}\amp(s,0) = s \sqrt{1-4m^2/s} \, \sigma_{\rm T}(s)$, where $\sigma_{\rm T}$ is the total cross section from $1 \, 2 \to {\rm everything}$.
In addition, unitarity and causality in theories with a mass gap imply that amplitudes are polynomially bounded as $\amp(s,t)/s^2 \to 0$ for $|s| \to \infty$ as a result of the Froissart-Jin-Martin bound. Even though here we are interested in theories with a massless graviton, it has been argued from different perspectives that a growth smaller than $s^2$ holds as well with dynamical gravity, see e.g.~\cite{Camanho:2014apa,Arkani-Hamed:2020blm,Chandorkar:2021viw,Caron-Huot:2021rmr,Haring:2022cyf}.
Note then that when this is the case, the integral \Eq{eq:dispersion} over a contour $\Gamma_\infty$ at $|s| \to \infty$ vanishes for $n \geqslant 2$, i.e.~$\Sigma_{n \geqslant 2} (\infty,t) = 0$.
A dispersion relation is then finally derived by using Cauchy's theorem to deform the original contour $\Gamma_0$ to $\Gamma_\infty$, leaving $\Sigma_n$ as an integral over the aforementioned singularities.
In the forward limit $t = 0$ of 4-scalar scattering, one then finds for $n=2$
\beq
c_2 = {\sum}_X \dfrac{2}{\pi} \int_0^\infty \dfrac{ds}{s^2} \sigma_{1_\phi 2_\phi \to X}(s) > 0 \, . \quad (\textrm{w/o GR's minimal coupling})
\label{eq:twicesub}
\eeq
This positivity constraint can be improved by noticing that, while the cross sections for production of the heavy states associated with the UV completion are by construction not computable within the EFT, those for production of the low-energy states are, as long as we restrict them to energies $s \leqslant \Lambda^2$ \cite{Nicolis:2009qm,Bellazzini:2016xrt,deRham:2017imi,Bellazzini:2017fep,deRham:2017xox}.
Since we are neglecting the minimal coupling of gravitons, the process with the largest cross section in the scalar EFT \Eq{eq:Shair} is the production of a pair of gravitons via the GB/CS coupling. The corresponding amplitude is
\beq
\amp^{\GB/\CS}_{1_\phi 2_\phi 3_{h^{++}} 4_{h^{--}}} = \left(\frac{2 |\hat{\alpha}|}{\mpl}\right)^2 \langle 4 | 1 | 3]^4 \left(\frac{1}{t} + \frac{1}{u} \right) \, .
\label{eq:phiphihhGB}
\eeq
Explicitly including this contribution in the twice-subtracted dispersion relation \Eq{eq:twicesub}, we arrive at
\beq
c_2 > \frac{2}{\pi} \int^{\Lambda^2}_0 \frac{ds}{s^2} \sigma^{\GB/\CS}_{\phi \phi \to h^{--} h^{++}} = 
\frac{1}{60 \pi^2} \left( \dfrac{|\hat{\alpha}| \Lambda^2}{\mpl} \right)^4 \, . \quad (\textrm{w/o GR's minimal coupling})
\label{eq:dispc2}
\eeq
If one could ignore GR's contributions to the dispersion relation, as we have done this far, such a beyond-positivity bound would imply that the GB/CS scalar-tensor theories in \Eq{eq:Shair} are inconsistent with unitarity and causality unless they are supplemented with the $(\nabla \phi)^4$ operator. In particular, note that the larger the regime of validity of the EFT, i.e.~the larger the cutoff $\Lambda$, the larger its coefficient $c_2$ would have to be.%
\footnote{In \Sec{sec:pheno} we discuss the effects of the leading additional operators in \Eq{eq:Sadd} on black holes with scalar hair. From that analysis one can arrive at the conclusion that for astrophysical black holes where $r_s  \sim |\hat{\alpha}|^{1/2} \sim \km$, the effects of $(\nabla \phi)^4$ with $c_2$ fixed by \Eq{eq:dispc2} would become $O(1)$, thus as important as the GB/CS term, for $\Lambda \gtrsim \mum^{-1}$, precisely of the same order as the smallest scales where gravity has been experimentally tested \cite{Lee:2020zjt} and at least up to which one would want any BGR theory to hold.}
However, neglecting GR's interactions would require in practice the existence of a consistent decoupling limit in which $\mpl \to \infty$ yet the lower bound on $c_2$ remains non-zero. Expressing \Eq{eq:dispc2} in terms of the strong coupling scale \Eq{eq:Lalpha}, $c_2 \gtrsim \tfrac{1}{16\pi^2}(\Lambda/\La)^8 \La^{-4}$, this would require keeping $\La$ as well as $\Lambda$ fixed. 
However, precisely because of the causality bound we derived in \Sec{sec:timeadv}, $\Lambda \lesssim 1/|\hat{\alpha}|^{1/2}$ for $\log(b_0 \Lambda) \sim 1$ (or equivalently $\Lambda \lesssim (\La^3/\mpl)^{1/2}$), such a limit is not possible: if $\mpl \to \infty$, then either $\La \to \infty$ or $\Lambda \to 0$, rendering the EFT invalid. 
In fact, even if one saturates the upper bound on $|\hat{\alpha}|$, the beyond-positivity contribution to $c_2$ in \Eq{eq:dispc2} is only as large as a quantum correction in GR at one loop, i.e.~$c_2 \gtrsim \tfrac{1}{16\pi^2}\mpl^{-4}$. 

We can explicitly check that one cannot ignore GR's minimal coupling if the upper bound $\Lambda \lesssim 1/|\hat{\alpha}|^{1/2}$ holds by retaking the steps above keeping $t \neq 0$ and with the low-energy contour now enclosing the graviton pole, at $s = 0$ and $s = -t$ ($u = 0$), of the 4-scalar amplitude in GR,
\beq
\amp^\GR_{1_\phi 2_\phi 3_\phi 4_\phi} = - \frac{1}{2 \mpl^2} \left( \frac{t^2 + u^2}{s} + \frac{s^2 + u^2}{t} + \frac{t^2 + s^2}{u} \right) \, .
\label{eq:phi4GR}
\eeq
Note that the forward limit is ill-defined because of the graviton $t$-channel exchange. The $n=2$ dispersion relation then reads
\beq
-\dfrac{1}{\mpl^2 t} + c_2 + \beta_2^{(t)} \log\dfrac{t}{t_0} + O(t)  = \dfrac{2}{\pi} \int_0^\infty ds \frac{{\rm Im}\amp(s',t)}{\left(s+\frac{t}{2}\right)^3} \, .
\label{eq:twicesubtimp}
\eeq
We have included the one-loop UV divergence of the $s^2$ term of the 4-scalar amplitude arising from $t$-channel cuts, with $\beta$-function given by $\beta_2^{(t)} = + (13/160 \pi^2) \mpl^{-4}$. This is of the same loop order as the r.h.s.~of \Eq{eq:dispc2}. Indeed, as discussed in \cite{Bellazzini:2020cot}, the beyond-positivity contributions to the dispersion relation are equivalent to including the running of the coefficients of the EFT in the forward limit, associated with the UV divergences from $s$- and $u$-channel cuts. These cuts and the corresponding gravitational $\beta$-functions can be easily computed following the on-shell amplitude techniques presented in \cite{Baratella:2021guc}.
The $O(t)$ term in \Eq{eq:twicesubtimp} encodes the subleading terms in the forward limit, arising from e.g.~higher-order EFT operators in the derivative expansion. For instance, the first such correction comes from the Galileon-like term $(\nabla \phi)^2 (\nabla \nabla \phi)^2$, which gives rise to an $stu$ term in the amplitude.
Most importantly, as advanced at the end of \Sec{sec:ndabounds}, the $1/t$ term in \Eq{eq:twicesubtimp} precludes setting a positive lower bound on $c_2$ as the one in \Eq{eq:dispc2}, unless the beyond-positivity contributions are larger than $-(\mpl^2 t)^{-1} \gtrsim (\mpl \Lambda)^{-2}$ \cite{Caron-Huot:2021rmr}. As we discussed above, this is not the case because of the causality bound $|\hat{\alpha}| \lesssim 1/\Lambda^2$.

While it might naively seem from the discussion above that the main obstruction for the derivation of meaningful beyond-positivity bounds in gravitational EFTs is the $t$-channel graviton pole, the real reason for their ineffectiveness is the fact that BGR amplitudes larger than in GR are not consistent with causality.
To show this, let us consider a dispersion relation for the 4-graviton amplitude with two positive and two negative helicities. The amplitude in GR plus the leading BGR correction in the energy expansion from \Eq{eq:Sadd} is given by
\beq
\amp^{\GR+\Delta S}_{1_{h^{++}} 2_{h^{--}} 3_{h^{--}} 4_{h^{++}}} (s,t) = \dfrac{\langle 2 3 \rangle^4 [1 4]^4}{\mpl^2} f (s,t) \, , \quad f (s,t) = \dfrac{1}{stu} + 8 (\alpha_4 + \alpha'_4) \,.
\label{eq:grav4alpha}
\eeq
Similarly to the scalar case, one can construct dispersion relations from the contour integral (see e.g.~\cite{Bellazzini:2015cra,Arkani-Hamed:2020blm,Bern:2021ppb,Caron-Huot:2022ugt} for more details)
\beq
\frac{1}{2 \pi i} \oint_{\Gamma_s} ds' \frac{f (s',t)}{\left(s'+\frac{t}{2}\right)^{n+1}} \, .
\eeq
In particular, for $n = 0$ one arrives at
\beq
\dfrac{8 (\alpha_4 + \alpha'_4)}{\mpl^2} + \dfrac{\gamma_4}{\Lambda^2 t} \log \dfrac{-t}{\mu^2} + O(t)  > \dfrac{2}{\pi} \int_0^{\Lambda^2} \frac{ds}{s^4} \sigma^{\GB/\CS}_{h^{++} h^{--} \to \phi \phi, \, h^{--} h^{++}} \sim \frac{1}{16 \pi^2} \left( \dfrac{|\hat{\alpha}| \Lambda}{\mpl} \right)^4 \, .
\label{eq:twicesubt}
\eeq
On the r.h.s.~we have explicitly included the beyond-positivity contribution from a scalar as well as a graviton loop via the GB/CS coupling, computed in a dispersive way from the corresponding cross sections. The corresponding amplitudes are given by \Eq{eq:phiphihhGB} and by
\beq
\amp^{\GB/\CS}_{1_{h^{++}} 2_{h^{--}} 3_{h^{++}} 4_{h^{--}}} = -\left(\frac{2 |\hat{\alpha}|}{\mpl}\right)^2 \dfrac{\langle 2 4 \rangle^4 [1 3]^4}{t} \, ,
\label{eq:hhhhGB}
\eeq
which proceeds via scalar exchange.
While there is no contribution from tree-level graviton exchange in \Eq{eq:twicesubt}, further forward limit singularities are generated at one loop  in GR, with $\gamma_4 \sim + (1/16 \pi^2) \mpl^{-4}$ \cite{Bern:2021ppb}.
Therefore, since the time-delay constraint $|\hat{\alpha}| \lesssim 1/\Lambda^2$ sets an upper bound on the r.h.s.~$\lesssim \tfrac{1}{16\pi^2} (\Lambda \mpl)^{-4}$, the beyond-positivity contribution is no larger than the one of GR, rendering the former immaterial to bound the quartic curvature operators.

In summary, because causality demands that gravitational amplitudes within the EFT domain are dominated by GR, loop corrections from BGR interactions never lead to robust lower bounds on the coefficients of the EFT.

\subsection{Power counting expectations}
\label{sec:ndaexpect}

The results of the previous section can be understood in terms of NDA, when the power counting rules in \Eq{eq:NDA} are extended to include the possibility that EFT operators can be generated at one-loop order,
\beq
\Lag = \tfrac{1}{2} \hat{M}_{\rm Pl}^2 R + \frac{\Lambda^4}{g^2} \left[ 
L^{(0)} \left( \frac{\nabla_\mu}{\Lambda}, \frac{R_{\mu \nu \rho \sigma}}{\Lambda^2}, \frac{g\phi}{\Lambda} \right)
+ \frac{g^2}{(4\pi)^2} L^{(1)} \left( \frac{\nabla_\mu}{\Lambda}, \frac{R_{\mu \nu \rho \sigma}}{\Lambda^2}, \frac{g\phi}{\Lambda} \right) + \cdots \right] 
\, .
\label{eq:NDAloop}
\eeq
This is because beyond-positivity contributions correspond to loop corrections within the EFT \cite{Bellazzini:2020cot,Arkani-Hamed:2020blm,Baratella:2021guc}. The one-loop NDA  estimate for the $(\nabla \phi)^4$ operator is 
$(c_2)^{(1)} \sim \tfrac{1}{16\pi^2} g^4 \Lambda^{-4}$, which for $g \sim \Lambda/\mpl$ matches the maximal value of the r.h.s.~of \Eq{eq:dispc2}, i.e.~for $|\hat{\alpha}| \sim 1/\Lambda^2$.
Likewise, we can estimate the beyond-positivity contributions to quartic curvatures operators from $L^{(1)}$ in \Eq{eq:NDAloop}, $(\alpha_4)^{(1)}/\mpl^2 \sim \tfrac{1}{16\pi^2} (\Lambda \mpl)^{-4}$, which coincides with the r.h.s.~of \Eq{eq:twicesubt} for the maximum value of the GB/CS coupling.

This discussion brings us to the important realization that, from the EFT standpoint, for UV completions where both the scalar-GB/CS term in \Eq{eq:Shair} and the operators in \Eq{eq:Sadd} are generated at the same (tree-level) order, one should expect much larger coefficients for the latter than what discussed above.
To see this, let us simply fix the power counting from the GB/CS term, assuming the maximal regime of validity of the EFT, $|\hat{\alpha}| \sim 1/\Lambda^2$ (a requirement, rather than a choice, if one interested in phenomenological applications, see \Sec{sec:pheno}).
From $L^{(0)}$ in \Eq{eq:NDAloop}, this sets $g \sim 1/|\hat{\alpha}| \Lambda \mpl \sim \Lambda/\mpl$, corresponding to a bona-fide UV completion of gravity, as discussed in \Sec{sec:ndabounds}.
Then, generic EFTs will feature
\beq
\alpha (\tilde{\alpha}) \sim \dfrac{1}{\Lambda^2} \,, \quad \alpha_3 \sim \dfrac{1}{\Lambda^4} \,, \quad \alpha_4, \alpha'_4 \sim \dfrac{1}{\Lambda^6} \,, \quad c_2 \sim \dfrac{1}{\Lambda^2\mpl^2} \,, \quad d_1 \sim \dfrac{1}{\Lambda^4} \, ,
\label{eq:NDASadd}
\eeq
for the coefficients of the operators in \Eq{eq:Sadd}.

In the next section we investigate the phenomenological consequences of these estimates for the physics of black holes with scalar hair.

\section{Phenomenological implications}
\label{sec:pheno}

In this section we discuss the main implications on the phenomenology of black holes of the upper bound on the GB/CS coupling $|\hat{\alpha}| \lesssim 1/\Lambda^2$, along with the implications associated with the additional EFT corrections that are expected from saturating such a bound, \Eq{eq:NDASadd}.
Our focus is on astrophysical black holes, in particular those detectable by LIGO-Virgo, which have sizes of a few solar masses, corresponding to Schwarzschild radii $r_s \gtrsim 10 \km$.

We will discuss separately the scalar-GB and dynamical-CS gravity theories, reviewing in each case their  imprints on the physics of black holes as well as the current experimental bounds on the couplings $\alpha$ and $\tilde{\alpha}$, respectively.

\subsection{Black holes in scalar-GB gravity}
\label{sec:GBpheno}

From a perturbative point of view, one can argue that the metric of a hairy black hole still displays a horizon, characterized by a linear zero of the metric, much like in the Schwarzschild and Kerr cases. The effects due to the BGR dynamics rapidly vanish far away from the horizon ($r = r_s$), following the fall-off of the GB invariant, which sources both the scalar hair and the deviations from GR in the metric. 

More in detail, in the static and spherically symmetric case, we have the metric
\beq
ds^2= -h(r)dt^2+f(r)^{-1}dr^2+r^2(d\theta^2+\sin^2\theta d\varphi^2) \,,
\eeq
with
\beq
h(r), f(r) \sim \left (1-\dfrac{r_s}{r}\right) \quad \mathrm{for} \quad r \sim r_s \,.
\eeq
At leading order in the dimensionless expansion parameter $\alpha/r^2$, the GB invariant is the one of the Schwarzschild solution,
\beq
\mathcal{R}_{\GB}^2 \equiv R^{\mu\nu\rho\sigma}R_{\mu\nu\rho\sigma}-4R^{\mu\nu}R_{\mu\nu}+R^2
\sim \dfrac{r_s^2}{r^6} \,.
\label{eq:GB-schw}
\eeq
The scalar equation of motion reads
\beq
\Box\phi = \mpl \alpha \mathcal{R}_{\GB}^2 \sim
\dfrac{\mpl^2}{\La^3} \dfrac{r_s^2}{r^6} \,,
\label{eq:eomscal}
\eeq
where in the last step we have traded the GB coupling $\alpha$ for the strong coupling scale $\La$, given in \Eq{eq:Lalpha} ($\tilde{\alpha} = 0$).
The scalar field profile is then completely determined by requiring that invariant quantities built out of it do not diverge for $r \geq r_s$ \cite{Mignemi:1992nt}.
At asymptotically large distances, $r \to \infty$, the solution behaves as $\phi(r) \sim 1/r$.
In addition, let us note that the largest value of the scalar radial derivative is estimated as,
\beq
\phi' \lesssim 
\frac{\mpl^2}{(\La r_s)^3} \,.
\label{eq:max-phi-prime}
\eeq
A first bound on the GB coupling comes from the requirement of the existence of real solutions for the scalar profile. In the simple EFT \Eq{eq:Shair}, this condition requires that $\alpha^2 < r_s^4/192$ \cite{Sotiriou:2014pfa}. The constraint is however dependent on additional EFT corrections from \Eq{eq:Sadd} \cite{Noller:2019chl}.

In order to estimate the impact of the scalar-GB operator on the background geometry, we can compute its ratio to GR, that is to $\mpl^2 \mathcal{R}$ where $\mathcal{R} = \sqrt{R_{\mu \nu \rho \sigma} R^{\mu \nu \rho \sigma}}$ is the typical curvature. Evaluating both terms on the background given by the Schwarzschild metric, $\mathcal{R} \sim r_s/r^3$, and the scalar solution from \Eq{eq:eomscal}, one finds \cite{Noller:2019chl}
\begin{equation}
\varepsilon_0(r) = \frac{\mpl \alpha  \phi \mathcal{R}^2_\GB}{\mpl^2 \mathcal{R}} \sim \left( \dfrac{\alpha}{r^2} \right)^2 \,.
\label{eq:e0}
\end{equation}

Turning to perturbations, let us start by noting that the theory \Eq{eq:Shair} has no scalar self-interactions. Therefore, there is simply no possible screening effect associated to classical non-linearities. On the other hand, there is no direct coupling between the scalar field and matter, therefore no screening mechanism is required to have agreement with fifth-force constraints (if the theory is valid at the scales of those experiments).

Instead, the scalar-GB term gives rise to a kinetic mixing between scalar and graviton (the same leading to the causality bound of \Sec{sec:scalartensor}), schematically of the form
\begin{equation}
\mpl \alpha \phi \mathcal{R}_\GB^2 \supset \varepsilon_{\rm mix}(r) \, \partial \phi \partial h \,,
\end{equation}
where we are taking the fluctuations to be canonically normalized. This effect will be important when the mixing
\begin{equation}
\varepsilon_{\rm mix}(r) \sim \dfrac{\alpha r_s}{r^3} \,,
\label{eq:emix}
\end{equation}
becomes of order one. Note that the two estimators of the BGR effects are related to each other, namely
\begin{equation}
\varepsilon_0 \sim \left( \dfrac{r}{r_s} \varepsilon_{\rm mix} \right)^2 \,.
\label{eq:bg-mixing-rel}
\end{equation}
In this scenario, a sizeable deviation of the quasi-normal mode (QNM) spectrum from the GR prediction is expected, strongly affecting the waveform during the ringdown phase of a merger.

Finally, let us turn to the phenomenology of a binary system of hairy black holes, each sourcing its own scalar profile as discussed before. The dynamical nature of the system implies that, just as it happens with gravitational waves, there will also be scalar wave emission. However, the latter is now dipolar instead of quadrupolar, therefore being much less suppressed than the former. This opens a new channel of power loss during the merger, which accelerates the rate of change in the orbital period. The effect acumulates during the inspiral phase, potentially producing an observable dephasing between the measured waveform and the one predicted by GR. To date, the absence of any observed effect of this type constitutes the most stringent experimental bound on the size of the scalar-GB coupling, $\alpha \lesssim (1.2 \km)^2$ \cite{Lyu:2022gdr}.
In terms of the strong coupling scale, this translates into $\La \gtrsim10^{12} \km^{-1}$. 
Let us add that such a bound strictly applies only to scalar-GB gravity with all other EFT corrections, in particular those in \Eq{eq:Sadd}, neglected or irrelevantly small.

Considering a typical LIGO/Virgo black hole, with $r_s \sim 10 \km$, the above bound implies that the kinetic mixing is constrained to be $\varepsilon_{\rm mix} \lesssim 10^{-2}$. Furthermore, according to \Eq{eq:bg-mixing-rel} the effect on the background geometry is significantly suppressed, $\varepsilon_0 \lesssim 10^{-4}$. This conclusion justifies neglecting deviations from Schwarzschild background as we assumed initially.

\subsection{EFT implications on scalar-GB black holes}
\label{sec:EFTpheno}

The fact that the BGR effects on hairy black holes are relatively small, below the $10\,\%$ level, could have been anticipated from the requirement that the scalar-GB theory should be able to properly describe the black holes of interest.
Indeed, as derived in \Sec{sec:scalartensor}, causality sets an upper bound on the GB coupling and therefore on the size of the observable corrections relative to GR, which we denote generically with $\epsilon(r)$ -- one instance being $\varepsilon_{\rm mix}$ in \Eq{eq:emix}. The BGR effects are largest near the horizon, 
\beq
\epsilon(r_s) \sim \dfrac{\alpha}{r_s^2} \lesssim (\Lambda r_s)^{-2} \,,
\label{eq:BGR}
\eeq
where the inequality follows from causality, \Eq{eq:boundGB} (with $\log(b_0\Lambda) \sim 1$ and $\tilde{\alpha} = 0$).
Usefulness of the EFT requires a hierarchy between the cutoff and the relevant scales of the system. 
A sensible demand on the EFT is therefore that the black hole falls within the EFT regime of validity at least down to its Schwarzschild radius.%
\footnote{As a matter of fact, one would like a gravitational EFT to be valid at least up to the scales where gravity has been experimentally tested, that is a cutoff $\Lambda \gtrsim \mum^{-1}$ \cite{Lee:2020zjt}. One is forced to give up on such a requirement if the scalar-GB theory is to be phenomenologically interesting for astrophysical black holes (see however the discussion in \Sec{conclusions}).}
Therefore, BGR corrections can never become large, i.e.~$\epsilon(r_s) \ll 1$.
Furthermore, taking the current experimental upper bound on $\alpha$ as benchmark, the causality bound implies a very low maximal cutoff,
\beq
\Lambda \lesssim (1 \km)^{-1} \, ,
\label{eq:maxcutoff}
\eeq
certainly much smaller than the strong coupling scale $\La \gtrsim (10^{-12} \km)^{-1}$. 

Let us discuss now the additional BGR effects that could be expected from the UV completion in the form of higher-dimensional operators with coefficients fixed to \Eq{eq:NDASadd}, where let us recall that such NDA estimates correspond to the the maximal cutoff $\Lambda \sim \alpha^{-1/2} \sim (1 \km)^{-1}$. 

The operators in \Eq{eq:Sadd} give rise to modifications of the geometry. We can estimate such modifications as in \Eq{eq:e0} for the scalar-GB term, which we recall scales as $\varepsilon_0(r) \sim (\alpha/r^2)^2 \sim (\Lambda r)^{-4}$. Similarly, we find
\beq
\!\!
\frac{\alpha_3 \mathcal{I}}{\mathcal{R}} \sim \dfrac{r_s^2}{r^2} (\Lambda r)^{-4} \, , 
\quad \!\!
\frac{\alpha_4 \mathcal{C}^2}{\mathcal{R}} \sim \dfrac{r_s^3}{r^3} (\Lambda r)^{-6} \, ,
\quad \!\!
\frac{c_2 (\nabla \phi)^4}{\mpl^2 \mathcal{R}} \sim \dfrac{r^5}{r_s^5} (\Lambda r)^{-10} \, ,
\quad \!\!
\frac{d_1 \mathcal{C} (\nabla \phi)^2}{\mpl^2 \mathcal{R}} \sim \dfrac{r}{r_s} (\Lambda r)^{-8} \, .
\eeq
While deviations introduced by the scalar-GB term are the largest, operators cubic in the Riemann tensor can become as important near the horizon. The deviations introduced by the rest of operators are subleading, being higher order in $(\Lambda r_s)^{-1}$, as expected from the derivative expansion and \Eq{eq:max-phi-prime}.
In addition, the operators built out of the scalar give rise to modifications of the scalar field profile, which we can estimate as
\beq
\frac{c_2 (\nabla \phi)^4}{(\nabla \phi)^2} \sim \dfrac{r^2}{r_s^2} (\Lambda r)^{-6} \, , 
\quad
\frac{d_1 \mathcal{C} (\nabla \phi)^2}{(\nabla \phi)^2} \sim \dfrac{r_s^2}{r^2} (\Lambda r)^{-4} \, . 
\eeq
Given the upper bound \Eq{eq:maxcutoff}, we conclude that the operators in \Eq{eq:Sadd} should induce corrections on the metric and scalar of up to $0.01\,\%$ near the horizon of black holes with $r_s \sim 10 \km$.

There are many other interesting signatures associated with the operators that we expect to be present in the scalar-GB EFT. The phenomenology of cubic and quartic curvature operators have been discussed in \cite{deRham:2020ejn,AccettulliHuber:2020dal,Cano:2019ore,Cano:2021myl,Silva:2022srr} and \cite{Endlich:2017tqa,Cardoso:2018ptl,Sennett:2019bpc,Silva:2022srr}, respectively.
Besides modifications of the Schwarzschild (and Kerr) geometries, these include deviations from GR at the leading order in QNMs, quadrupole moments, non-vanishing Love numbers, and corrections to the gravitational-wave signals at relatively high post-Newtonian (PN) order. 
Since such corrections start at order $(\Lambda r)^{-4} \lesssim 10^{-4}$, they will not be easy to probe with the sensitivity of current experiments.

Operators involving the scalar field have received less attention in the literature. The impact of the cubic Galileon operator, which we have rewritten in \Eq{eq:Sadd} as $d_1 R_{\mu \nu \rho \sigma}^2 (\nabla_\eta \phi)^2$, has been discussed in \cite{Noller:2019chl}. Along with the operator $c_2 (\nabla_\mu \phi)^4$, these EFT terms introduce modifications e.g.~in the scalar and gravitational wave spectrum, as well as in the QNMs, predicted by the pure scalar-GB theory \Eq{eq:Shair}, although a priori subleading due to the suppression by higher powers of $(\Lambda r)^{-2} \ll 1$. In particular, note that potential screening effects are not likely to be significant.

Let us recall once again that these conclusions appear to be a robust consequence of causality. Nevertheless, quantitative predictions for the gravitational observables, which typically require performing costly numerical simulations, are still interesting, if only to experimentally test the fundamental principles behind these expectations.

\subsection{Black holes in dynamical-CS gravity}

The pseudo-scalar-CS term in \Eq{eq:Shair} (with $\alpha = 0$) gives rise to a phenomenology of black holes similar to that discussed in the previous section, although with a few important differences.
First, the Pontryagin invariant, $R_{\mu\nu\rho\sigma}\tilde{R}^{\mu\nu\rho\sigma}$, vanishes in the Schwarzschild geometry, while it is non-zero for the Kerr geometry. Therefore, one needs to consider rotating black holes in order to have a non vanishing scalar hair. 
To simplify the analysis, we follow \cite{Yunes:2009hc} and treat the spin parameter of the black hole, $a/r_s$, perturbatively. We also work in an expansion in $\tilde{\alpha}/r_s^2$ since, similar to discussion for scalar-GB black holes in \Sec{sec:EFTpheno}, this is a consequence of causality, $\tilde{\alpha} \lesssim 1/\Lambda^2$, along with the requirement that the black holes of interest fall within the regime of validity of the EFT, i.e.~$(\Lambda r_s)^{-2} \ll 1$.

At leading order, the equation of motion for the pseudo-scalar $\phi$ is given by
\beq
\Box\phi = \mpl \tilde{\alpha} R_{\mu\nu\rho\sigma}\tilde{R}^{\mu\nu\rho\sigma} 
\sim \dfrac{\mpl^2}{\Lat^3} \dfrac{r_s^2}{r^6}\,\dfrac{a}{r} \cos\theta \,,
\label{eq:dCSeom}
\eeq
where in the last step we have traded the CS coupling $\tilde{\alpha}$ for the strong coupling scale $\Lat$, given in \Eq{eq:Lalpha} ($\alpha = 0$ and changed subscript to avoid confusion). Subleading terms in the spin parameter scale as $(a/r_s)^3$.
From the solution to this equation (see e.g.~\cite{Yunes:2009hc}), we can compute the scalar radial derivative, which roughly satisfies
\beq
\phi' \lesssim 
\frac{\mpl^2}{(\Lat r_s)^3} \dfrac{a}{r_s} \,.
\label{eq:max-phi-prime-CS}
\eeq
Similarly to scalar-GB case, there is an approximate relation between corrections to the Kerr geometry, $\varepsilon_0$, and the kinetic mixing between pseudo-scalar and graviton, $\varepsilon_{\rm mix}$, given by
\begin{equation}
\varepsilon_0(r) \sim \left( \dfrac{a}{r_s} \varepsilon_{\rm mix}(r) \right)^2 \,.
\end{equation}

A second difference w.r.t.~the scalar-GB case is the nature of the most stringent experimental bounds on the CS coupling. Since the scalar background sourced around an isolated (spinning) black hole is dipolar, the emission of scalar waves from a black hole binary system starts from a quadrupole moment. Therefore, energy loss via scalar emission is further suppressed in the PN expansion compared to the scalar-GB case, such that no constraint can be derived on $\tilde{\alpha}$ given current sensitivities \cite{Nair:2019iur}.

The strongest bound to date on the pseudo-scalar-CS coupling comes from independent measurements of the tidal deformability and of the moment of inertia in neutron stars \cite{Silva:2020acr}. The comparison between these measurements and the values predicted in dynamical-CS gravity yields the bound $\tilde{\alpha} \lesssim (8 \km)^2$. Note this is weaker than the most stringent bound on the scalar-GB coupling, $\alpha \lesssim (1.2 \km)^2 $ by one order of magnitude. Nevertheless, if the pseudo-scalar-CS EFT is to be able to describe black holes with sizes down to $r_s \sim 10 \km$ (recall that the smallest black holes display the largest BGR effects, which in any case cannot become $O(1)$) with at least $10\,\%$ accuracy, it seems wise to consider a different benchmark for the coupling $\tilde{\alpha}$. 
Maximal testability compatible with causality then suggests to take $\tilde{\alpha} \sim 1/\Lambda^2 \sim (3 \km)^2$.

In this case, the kinetic mixing between pseudo-scalar and graviton can induce stronger deviations in the QNM spectrum w.r.t.~to GR \cite{Cano:2020cao,Wagle:2021tam,Srivastava:2021imr}, of order $\varepsilon_{\rm mix} \lesssim (\tilde{\alpha}/r_s^2) \sim 10^{-1}$ for black holes with $r_s \sim 10 \km$.

The discussion of the implications of the additional operators in \Eq{eq:Sadd}, present in generic UV completions of pseudo-scalar-CS theory, largely parallels that of \Sec{sec:EFTpheno} and we do not repeat it here. Nevertheless, we wish to point out that to date much less work has been devoted to the study of these EFT effects for rotating black holes, see e.g.~\cite{Cano:2019ore,Cano:2021myl}.

Before closing the section, let us point out one last difference between the scalar-GB and dynamical-CS theories. While the scalar-GB operator leads to equations of motion of at most second order in (time) derivatives, the dynamical-CS operator gives rise to higher-derivative terms. These in principle could spoil the quantum stability of the theory. 
Considering perturbations of rotating black holes with pseudo-scalar-CS hair, for instance during the inspiral phase of a merger,%
\footnote{This system allows us to consider non-vanishing time derivatives of the scalar background, which can lead to ghosts instabilities.}
higher derivatives will become important at a mass scale $M_g^{-1} \sim \tilde{\alpha} \dot{\phi}_0/\mpl$, being $ \dot{\phi}_0\sim \omega \phi_0 \lesssim \phi_0/r_s$ the time derivative of the axionic field evaluated on the background. We can estimate this scale, at least in some appropriate regime, using the solution of \Eq{eq:dCSeom}, finding
\beq
M_g \sim \dfrac{1}{a} \left(\dfrac{r_s^2}{\tilde{\alpha}}\right)^2 \,.
\label{eq:massghost}
\eeq
Since causality requires $ \Lambda \lesssim \tilde{\alpha}^{-1/2} $, we find that the ghost's mass is above the EFT cutoff.

\section{Summary and Outlook}
\label{conclusions}
Gravitational wave science can potentially test deviations from General Relativity. While the EFT framework gives the best organizing principle to characterize physics beyond GR, the complexity of black hole merger events suggests that the theory space should first be reduced as much as possible before comparing data with predictions coming from different EFTs. A powerful and robust set of constraints on the space of consistent EFTs comes from the unitarity and causality of the (unknown) UV completion. 

In this paper, we used causality arguments to derive phenomenologically interesting bounds on theories that, at low energies, comprise the graviton and a shift-symmetric 
scalar field.  
The presence of a coupling between the scalar 
and the Gauss-Bonnet or the Chern-Simons operators leads to black hole hair. If the couplings $\alpha$ and $\tilde{\alpha}$, in the notation of \Eq{eq:Shair}, are large enough, $|\hat{\alpha}| \sim r_s^2$, hair can be measured in astrophysical black holes. The first consequence that we derived imposing the absence of a (resolvable) time advance is that the cutoff of the EFT cannot be parametrically larger than $|\hat{\alpha}|^{-1/2}$, i.e.~$1/\km$ for solar mass BHs.
This has implications for the structure of all the higher-dimensional operators in the theory. One could attempt to draw robust lower bounds on their coefficients using positivity constraints obtained via dispersion relations. However, the weakness of non-minimal gravitational interactions compared to GR, enforced by causality, implies that lower bounds from one-loop dispersion relations are phenomenologically irrelevant. 

On the other hand, for such a low cutoff $\Lambda \sim 1/\km$, we used general power counting arguments to show that if both the scalar-GB/CS term and other operators are generated at the same (tree-level) order, the latter will also give sizable contributions in black hole dynamics.
            
The result that the UV cutoff of a phenomenologically interesting and causal EFT describing gravity must be lower than $\invkm$ should not be considered only as a source of potentially large corrections to the effective description of astrophysical black holes. Instead, we should demand that this result is reconciled with our experimental knowledge of gravity.
As a matter of fact, to date gravity has been probed in table-top experiments down to the scales of tens of microns in length, showing good agreement with Newtonian theory \cite{Lee:2020zjt}. In light of this, one should at least prefer, if not demand, that GR is extended using a theory valid down to the $\mu$m, in such a way to describe the same observations as GR does.

In the scenarios studied above, being the cutoff at a much lower scale than $\mum^{-1}$, one needs to trust that the UV completion that gives rise either to the scalar-GB or to the dynamical-CS EFTs, does indeed reproduce the Newtonian potential at microscopic lengths. 
Similarly to what was argued in \cite{Endlich:2017tqa,Sennett:2019bpc} for quartic curvature terms, this might be the case of a ``soft'' UV completion that resolves the irrelevant operators in the EFT in such a way that interactions stop growing with the energy.

In addition to this requirement, we need the UV completion to contribute to the time delay in such a way that   causality is preserved, in particular given the negative contributions (i.e.~the time advance) from the low-energy operators. 

This requirement is a substantial obstacle for both the scalar-GB and dynamical-CS EFTs.%
\footnote{As a matter of fact, consistency with unitarity and causality is an obstacle as well for any BGR deformation of phenomenological relevance above the $\mu$m, in particular if it involves higher-order terms in the curvature \cite{Caron-Huot:2022ugt}.} Indeed, since the sizes of both the scalar-GB and of the axion-CS operators are chosen in such a way to follow the tree-level NDA, it appears difficult for loop-level effects in the UV physics to restore causality, unless the number of species scales as $(4 \pi/g)^2 \sim (4 \pi \mpl/\Lambda)^2$ \cite{Veneziano:2001ah,Dvali:2007hz}. On the other hand, as it was argued in \cite{Camanho:2014apa,Kologlu:2019bco}, in order for causality to be restored by tree-level exchanges in the UV, one must introduce an infinite tower of higher-spin particles having a mass of order $M \sim \invkm$. 
In both these cases, the description of gravity below the km would be very different from what we know. 


In light of this, there seem to be two reasonable attitudes. The first is to resign to the idea that both scalar-GB and pseudo-scalar-CS interactions cannot lead to testable modifications of GR. The second is trying to understand if there exists a UV completion of these models that restores causality without clashing with our knowledge of gravity at small distances.
In any case, the detection of black hole hair would be revolutionary, telling us there is something fundamentally unexpected and so far unknown about gravitational dynamics.


\section*{Acknowledgments}
J.S.~thanks Brando Bellazzini for illuminating discussions. J.S.~has been partially supported by the DFG Cluster of Excellence 2094 ORIGINS and the Collaborative Research Center SFB1258.
E.T.~is partly supported by the Italian MIUR under contract 2017FMJFMW (PRIN2017). The work of L.G.T.~is supported by the Grant Agency of the Czech Republic, GACR grant 20-28525S.

\appendix 

\section{Gauss-Bonnet scalarization}
\label{sec:scalarization}

For scalar-tensor theories with no scalar shift-symmetry, one can consider non-linear couplings between $\phi$ and the GB invariant. Let us focus on the leading $Z_2$-symmetric ($\phi \to - \phi$) term in a field and derivative expansion,
\beq
S_{\GBsc} = \int d^4 x \sqrt{-g} \left( \dfrac{\mpl^2}{2} R - \dfrac{1}{2} (\nabla_\mu \phi)^2 + \lambda \phi^2 \mathcal{R}_{\GB}^2 \right) \,,
\label{eq:SGBization}
\eeq
where $\lambda$ has dimensions of a length square. 
The sign of $\lambda$ determines whether $\phi = 0$ is a stable solution or not around a gravitational source like a (static or spinning) black hole \cite{Dima:2020yac}. Spontaneous scalarization, i.e.~a non-trivial scalar profile, generically develops for $|\lambda|/r_s^2 \gtrsim 1$, with $\lambda > 0$ provided the GB invariant is positive.\footnote{This is the case e.g.~on a static black hole background or away from the horizon on a rotating black hole background.}

From simple little group (helicity) selection rules, one can derive the 4-point interaction of two scalars and two gravitons associated with the non-minimal coupling to GB,
\beq
\amp^\GBsc_{1_\phi 2_\phi 3_{h^{++}} 4_{h^{++}}} = \frac{4 \lambda}{\mpl^2} [34]^4 \, , \quad \amp^\GBsc_{1_\phi 2_\phi 3_{h^{--}} 4_{h^{--}}} = \frac{4 \lambda}{\mpl^2} \langle 34 \rangle^4 \, .
\label{eq:4pointGBization}
\eeq
Note that this is associated with an inelastic $\phi h^{++} \to \phi h^{--}$ scattering amplitude, that vanishes in the forward limit $t \to 0$,
\beq
\amp^\GBsc_{1_\phi 2_{h^{++}} \to 3_\phi 4_{h^{--}}} = \frac{4 \lambda}{\mpl^2} t^2 e^{4i\theta} \, ,
\label{eq:phihphihGBization}
\eeq
where $\theta$ is just a phase (for physical momenta $[ij]^* = \langle ij \rangle = \sqrt{s_{ij}} e^{i \theta}$, with $s_{13} = t$).

While a dispersion relation for $\lambda$ from the 2-scalar-2-graviton amplitude can be derived along the lines of \Sec{sec:beyond}, the inelasticity of the amplitude preclude the derivation of a positivity bound $\lambda  > 0$.
One can actually come up with a simple (yet partial) tree-level UV completion that shows that the sign of $\lambda$ is not fixed.
This involves an additional massive scalar field $\Phi$,
\beq
S_{\GBsc}^\UV = \int d^4 x \sqrt{-g} \left( \tfrac{1}{2} \mpl^2 R - \tfrac{1}{2} (\nabla_\mu \phi)^2 - \tfrac{1}{2} (\nabla_\mu \Phi)^2 - \tfrac{1}{2} m_\Phi^2 \Phi^2 + \mpl \alpha_\Phi \Phi \mathcal{R}_{\GB}^2 + g_\Phi \Phi \phi^2 \right) \,.
\label{eq:UVSGBization}
\eeq
One can see that upon integrating out the massive scalar, one obtains
\beq
\lambda = \frac{\mpl \alpha_\Phi g_\Phi}{m_\Phi^2} \, ,
\label{eq:lambdaeff}
\eeq
with no definite sign, since e.g.~$\alpha_\Phi$ can consistently be positive or negative.
In addition, from this example one can infer that the size of $\lambda$ is likely to be theoretically bounded.
Indeed, since \Eq{eq:UVSGBization} is itself an effective action, from the generic EFT perspective presented in \Sec{sec:ndabounds}, the trilinear coupling is of order $g_\Phi \sim g \Lambda$ while $\alpha_\phi \sim 1/(g \mpl \Lambda)$, where $\Lambda$ is the cutoff and $g$ a coupling.
The mass of the heavy scalar can be at most of the order of the cutoff, i.e.~$m_\Phi \sim \Lambda$, and should itself be identified with the cutoff of \Eq{eq:SGBization}. We are then led to the conjecture that the quadratic scalar-GB coupling should not be much larger than $\lambda \sim 1/\Lambda^2$ given the causality bound on $\alpha_\Phi$.
Following similar arguments, one can start with the effective interaction $\lambda \phi^2 \mathcal{R}^2_{\GB}$ and give the scalar a vacuum expectation value, which within the EFT can be at most $\vev{\phi} \sim \Lambda/g \lesssim \mpl$. This gives rise to the scalar-GB term in \Eq{eq:Shair} with $\mpl \alpha \sim \lambda \vev{\phi}$, which for $\lambda \lesssim 1/\Lambda^2$ is consistent with the causality bound we derived in \Sec{sec:scalartensor}. Note that, at the end of the day, these arguments are just refined versions of the statement that, from the gravitational power counting discussed in \Sec{sec:ndabounds}, we expect $\lambda \sim \zeta^2/\Lambda^2 \lesssim 1/\Lambda^2$ for $\zeta \lesssim 1$.
However, in this case one cannot reinterpret such expectation as a consequence of the requirement that the BGR amplitude, \Eq{eq:4pointGBization}, should not become larger than GR's within the EFT, since the latter vanishes at tree level (and at one loop) \cite{Baratella:2021guc}.

Besides, one can construct a dispersion relation for the 4-scalar amplitude along the lines of \Sec{sec:beyond}. In this case, the beyond positivity bound corresponding to \Eq{eq:SGBization} is associated with a cross section for $\phi \phi \to h^{\pm\pm}h^{\pm\pm}$ which, similar to \Eq{eq:dispc2}, leads to $c_2 \gtrsim \tfrac{1}{16 \pi^2} \lambda^2 (\Lambda/\mpl)^4$.
If indeed $\lambda \lesssim 1/\Lambda^2$ regardless of the UV completion as long as this is unitary and causal, then the lower bound on $c_2$ is in fact inapplicable due to the graviton pole.

Finally, from the phenomenological point of view, scalarization of black holes turns out to be a dubious phenomenon, given that  $\lambda/r_s^2 \lesssim (\Lambda r_s)^{-2} \lesssim 1$ if the EFT is to describe the black holes down to their horizon.
However, let us recall that, differently from the cases discussed in the main text, for the theory of GB-scalarization we have not found solid evidence that causality forces the BGR effects to be subleading.

\bibliographystyle{JHEP.bst}
\bibliography{causalityGB}

\providecommand{\href}[2]{#2}\begingroup\raggedright\begin{thebibliography}{10}

\bibitem{Sotiriou:2013qea}
T.P.~Sotiriou and S.-Y.~Zhou, \emph{{Black hole hair in generalized
  scalar-tensor gravity}},
  \href{https://doi.org/10.1103/PhysRevLett.112.251102}{\emph{Phys. Rev. Lett.}
  {\bfseries 112} (2014) 251102}
  [\href{https://arxiv.org/abs/1312.3622}{{\ttfamily 1312.3622}}].

\bibitem{Sotiriou:2014pfa}
T.P.~Sotiriou and S.-Y.~Zhou, \emph{{Black hole hair in generalized
  scalar-tensor gravity: An explicit example}},
  \href{https://doi.org/10.1103/PhysRevD.90.124063}{\emph{Phys. Rev. D}
  {\bfseries 90} (2014) 124063}
  [\href{https://arxiv.org/abs/1408.1698}{{\ttfamily 1408.1698}}].

\bibitem{Yunes:2009hc}
N.~Yunes and F.~Pretorius, \emph{{Dynamical Chern-Simons Modified Gravity. I.
  Spinning Black Holes in the Slow-Rotation Approximation}},
  \href{https://doi.org/10.1103/PhysRevD.79.084043}{\emph{Phys. Rev. D}
  {\bfseries 79} (2009) 084043}
  [\href{https://arxiv.org/abs/0902.4669}{{\ttfamily 0902.4669}}].

\bibitem{Hui:2012qt}
L.~Hui and A.~Nicolis, \emph{{No-Hair Theorem for the Galileon}},
  \href{https://doi.org/10.1103/PhysRevLett.110.241104}{\emph{Phys. Rev. Lett.}
  {\bfseries 110} (2013) 241104}
  [\href{https://arxiv.org/abs/1202.1296}{{\ttfamily 1202.1296}}].

\bibitem{Creminelli:2020lxn}
P.~Creminelli, N.~Loayza, F.~Serra, E.~Trincherini and L.G.~Trombetta,
  \emph{{Hairy Black-holes in Shift-symmetric Theories}},
  \href{https://doi.org/10.1007/JHEP08(2020)045}{\emph{JHEP} {\bfseries 08}
  (2020) 045} [\href{https://arxiv.org/abs/2004.02893}{{\ttfamily
  2004.02893}}].

\bibitem{Herdeiro:2015waa}
C.A.R.~Herdeiro and E.~Radu, \emph{{Asymptotically flat black holes with scalar
  hair: a review}}, \href{https://doi.org/10.1142/S0218271815420146}{\emph{Int.
  J. Mod. Phys. D} {\bfseries 24} (2015) 1542014}
  [\href{https://arxiv.org/abs/1504.08209}{{\ttfamily 1504.08209}}].

\bibitem{Lee:2020zjt}
J.G.~Lee, E.G.~Adelberger, T.S.~Cook, S.M.~Fleischer and B.R.~Heckel,
  \emph{{New Test of the Gravitational $1/r^2$ Law at Separations down to 52
  $\mu$m}}, \href{https://doi.org/10.1103/PhysRevLett.124.101101}{\emph{Phys.
  Rev. Lett.} {\bfseries 124} (2020) 101101}
  [\href{https://arxiv.org/abs/2002.11761}{{\ttfamily 2002.11761}}].

\bibitem{Drummond:1979pp}
I.T.~Drummond and S.J.~Hathrell, \emph{{QED Vacuum Polarization in a Background
  Gravitational Field and Its Effect on the Velocity of Photons}},
  \href{https://doi.org/10.1103/PhysRevD.22.343}{\emph{Phys. Rev. D} {\bfseries
  22} (1980) 343}.

\bibitem{Camanho:2014apa}
X.O.~Camanho, J.D.~Edelstein, J.~Maldacena and A.~Zhiboedov, \emph{{Causality
  Constraints on Corrections to the Graviton Three-Point Coupling}},
  \href{https://doi.org/10.1007/JHEP02(2016)020}{\emph{JHEP} {\bfseries 02}
  (2016) 020} [\href{https://arxiv.org/abs/1407.5597}{{\ttfamily 1407.5597}}].

\bibitem{Bern:2021ppb}
Z.~Bern, D.~Kosmopoulos and A.~Zhiboedov, \emph{{Gravitational effective field
  theory islands, low-spin dominance, and the four-graviton amplitude}},
  \href{https://doi.org/10.1088/1751-8121/ac0e51}{\emph{J. Phys. A} {\bfseries
  54} (2021) 344002} [\href{https://arxiv.org/abs/2103.12728}{{\ttfamily
  2103.12728}}].

\bibitem{Caron-Huot:2022ugt}
S.~Caron-Huot, Y.-Z.~Li, J.~Parra-Martinez and D.~Simmons-Duffin,
  \emph{{Causality constraints on corrections to Einstein gravity}},
  \href{https://arxiv.org/abs/2201.06602}{{\ttfamily 2201.06602}}.

\bibitem{Silva:2017uqg}
H.O.~Silva, J.~Sakstein, L.~Gualtieri, T.P.~Sotiriou and E.~Berti,
  \emph{{Spontaneous scalarization of black holes and compact stars from a
  Gauss-Bonnet coupling}},
  \href{https://doi.org/10.1103/PhysRevLett.120.131104}{\emph{Phys. Rev. Lett.}
  {\bfseries 120} (2018) 131104}
  [\href{https://arxiv.org/abs/1711.02080}{{\ttfamily 1711.02080}}].

\bibitem{Macedo:2019sem}
C.F.B.~Macedo, J.~Sakstein, E.~Berti, L.~Gualtieri, H.O.~Silva and
  T.P.~Sotiriou, \emph{{Self-interactions and Spontaneous Black Hole
  Scalarization}},
  \href{https://doi.org/10.1103/PhysRevD.99.104041}{\emph{Phys. Rev. D}
  {\bfseries 99} (2019) 104041}
  [\href{https://arxiv.org/abs/1903.06784}{{\ttfamily 1903.06784}}].

\bibitem{AccettulliHuber:2020oou}
M.~Accettulli~Huber, A.~Brandhuber, S.~De~Angelis and G.~Travaglini,
  \emph{{Eikonal phase matrix, deflection angle and time delay in effective
  field theories of gravity}},
  \href{https://doi.org/10.1103/PhysRevD.102.046014}{\emph{Phys. Rev. D}
  {\bfseries 102} (2020) 046014}
  [\href{https://arxiv.org/abs/2006.02375}{{\ttfamily 2006.02375}}].

\bibitem{Gao:2000ga}
S.~Gao and R.M.~Wald, \emph{{Theorems on gravitational time delay and related
  issues}}, \href{https://doi.org/10.1088/0264-9381/17/24/305}{\emph{Class.
  Quant. Grav.} {\bfseries 17} (2000) 4999}
  [\href{https://arxiv.org/abs/gr-qc/0007021}{{\ttfamily gr-qc/0007021}}].

\bibitem{Edelstein:2021jyu}
J.D.~Edelstein, R.~Ghosh, A.~Laddha and S.~Sarkar, \emph{{Causality constraints
  in Quadratic Gravity}},
  \href{https://doi.org/10.1007/JHEP09(2021)150}{\emph{JHEP} {\bfseries 09}
  (2021) 150} [\href{https://arxiv.org/abs/2107.07424}{{\ttfamily
  2107.07424}}].

\bibitem{Goon:2016une}
G.~Goon and K.~Hinterbichler, \emph{{Superluminality, black holes and EFT}},
  \href{https://doi.org/10.1007/JHEP02(2017)134}{\emph{JHEP} {\bfseries 02}
  (2017) 134} [\href{https://arxiv.org/abs/1609.00723}{{\ttfamily
  1609.00723}}].

\bibitem{deRham:2019ctd}
C.~de~Rham and A.J.~Tolley, \emph{{Speed of gravity}},
  \href{https://doi.org/10.1103/PhysRevD.101.063518}{\emph{Phys. Rev. D}
  {\bfseries 101} (2020) 063518}
  [\href{https://arxiv.org/abs/1909.00881}{{\ttfamily 1909.00881}}].

\bibitem{deRham:2020zyh}
C.~de~Rham and A.J.~Tolley, \emph{{Causality in curved spacetimes: The speed of
  light and gravity}},
  \href{https://doi.org/10.1103/PhysRevD.102.084048}{\emph{Phys. Rev. D}
  {\bfseries 102} (2020) 084048}
  [\href{https://arxiv.org/abs/2007.01847}{{\ttfamily 2007.01847}}].

\bibitem{Bellazzini:2021shn}
B.~Bellazzini, G.~Isabella, M.~Lewandowski and F.~Sgarlata,
  \emph{{Gravitational Causality and the Self-Stress of Photons}},
  \href{https://arxiv.org/abs/2108.05896}{{\ttfamily 2108.05896}}.

\bibitem{Chen:2021bvg}
C.Y.R.~Chen, C.~de~Rham, A.~Margalit and A.J.~Tolley, \emph{{A cautionary case
  of casual causality}},  \href{https://arxiv.org/abs/2112.05031}{{\ttfamily
  2112.05031}}.

\bibitem{deRham:2021bll}
C.~de~Rham, A.J.~Tolley and J.~Zhang, \emph{{Causality Constraints on
  Gravitational Effective Field Theories}},
  \href{https://arxiv.org/abs/2112.05054}{{\ttfamily 2112.05054}}.

\bibitem{Amati:1992zb}
D.~Amati, M.~Ciafaloni and G.~Veneziano, \emph{{Planckian scattering beyond the
  semiclassical approximation}},
  \href{https://doi.org/10.1016/0370-2693(92)91366-H}{\emph{Phys. Lett. B}
  {\bfseries 289} (1992) 87}.

\bibitem{Kabat:1992tb}
D.N.~Kabat and M.~Ortiz, \emph{{Eikonal quantum gravity and Planckian
  scattering}}, \href{https://doi.org/10.1016/0550-3213(92)90627-N}{\emph{Nucl.
  Phys. B} {\bfseries 388} (1992) 570}
  [\href{https://arxiv.org/abs/hep-th/9203082}{{\ttfamily hep-th/9203082}}].

\bibitem{Akhoury:2013yua}
R.~Akhoury, R.~Saotome and G.~Sterman, \emph{{High Energy Scattering in
  Perturbative Quantum Gravity at Next to Leading Power}},
  \href{https://doi.org/10.1103/PhysRevD.103.064036}{\emph{Phys. Rev. D}
  {\bfseries 103} (2021) 064036}
  [\href{https://arxiv.org/abs/1308.5204}{{\ttfamily 1308.5204}}].

\bibitem{Bellazzini:2019xts}
B.~Bellazzini, M.~Lewandowski and J.~Serra, \emph{{Positivity of Amplitudes,
  Weak Gravity Conjecture, and Modified Gravity}},
  \href{https://doi.org/10.1103/PhysRevLett.123.251103}{\emph{Phys. Rev. Lett.}
  {\bfseries 123} (2019) 251103}
  [\href{https://arxiv.org/abs/1902.03250}{{\ttfamily 1902.03250}}].

\bibitem{Kologlu:2019bco}
M.~Kologlu, P.~Kravchuk, D.~Simmons-Duffin and A.~Zhiboedov, \emph{{Shocks,
  Superconvergence, and a Stringy Equivalence Principle}},
  \href{https://doi.org/10.1007/JHEP11(2020)096}{\emph{JHEP} {\bfseries 11}
  (2020) 096} [\href{https://arxiv.org/abs/1904.05905}{{\ttfamily
  1904.05905}}].

\bibitem{Veneziano:2001ah}
G.~Veneziano, \emph{{Large N bounds on, and compositeness limit of, gauge and
  gravitational interactions}},
  \href{https://doi.org/10.1088/1126-6708/2002/06/051}{\emph{JHEP} {\bfseries
  06} (2002) 051} [\href{https://arxiv.org/abs/hep-th/0110129}{{\ttfamily
  hep-th/0110129}}].

\bibitem{Dvali:2007hz}
G.~Dvali, \emph{{Black Holes and Large N Species Solution to the Hierarchy
  Problem}}, \href{https://doi.org/10.1002/prop.201000009}{\emph{Fortsch.
  Phys.} {\bfseries 58} (2010) 528}
  [\href{https://arxiv.org/abs/0706.2050}{{\ttfamily 0706.2050}}].

\bibitem{Arkani-Hamed:2006emk}
N.~Arkani-Hamed, L.~Motl, A.~Nicolis and C.~Vafa, \emph{{The String landscape,
  black holes and gravity as the weakest force}},
  \href{https://doi.org/10.1088/1126-6708/2007/06/060}{\emph{JHEP} {\bfseries
  06} (2007) 060} [\href{https://arxiv.org/abs/hep-th/0601001}{{\ttfamily
  hep-th/0601001}}].

\bibitem{Liu:2016idz}
D.~Liu, A.~Pomarol, R.~Rattazzi and F.~Riva, \emph{{Patterns of Strong Coupling
  for LHC Searches}},
  \href{https://doi.org/10.1007/JHEP11(2016)141}{\emph{JHEP} {\bfseries 11}
  (2016) 141} [\href{https://arxiv.org/abs/1603.03064}{{\ttfamily
  1603.03064}}].

\bibitem{Bellazzini:2015cra}
B.~Bellazzini, C.~Cheung and G.N.~Remmen, \emph{{Quantum Gravity Constraints
  from Unitarity and Analyticity}},
  \href{https://doi.org/10.1103/PhysRevD.93.064076}{\emph{Phys. Rev. D}
  {\bfseries 93} (2016) 064076}
  [\href{https://arxiv.org/abs/1509.00851}{{\ttfamily 1509.00851}}].

\bibitem{Hamada:2018dde}
Y.~Hamada, T.~Noumi and G.~Shiu, \emph{{Weak Gravity Conjecture from Unitarity
  and Causality}},
  \href{https://doi.org/10.1103/PhysRevLett.123.051601}{\emph{Phys. Rev. Lett.}
  {\bfseries 123} (2019) 051601}
  [\href{https://arxiv.org/abs/1810.03637}{{\ttfamily 1810.03637}}].

\bibitem{Chowdhury:2019kaq}
S.D.~Chowdhury, A.~Gadde, T.~Gopalka, I.~Halder, L.~Janagal and S.~Minwalla,
  \emph{{Classifying and constraining local four photon and four graviton
  S-matrices}}, \href{https://doi.org/10.1007/JHEP02(2020)114}{\emph{JHEP}
  {\bfseries 02} (2020) 114}
  [\href{https://arxiv.org/abs/1910.14392}{{\ttfamily 1910.14392}}].

\bibitem{Tokuda:2020mlf}
J.~Tokuda, K.~Aoki and S.~Hirano, \emph{{Gravitational positivity bounds}},
  \href{https://doi.org/10.1007/JHEP11(2020)054}{\emph{JHEP} {\bfseries 11}
  (2020) 054} [\href{https://arxiv.org/abs/2007.15009}{{\ttfamily
  2007.15009}}].

\bibitem{Alberte:2020bdz}
L.~Alberte, C.~de~Rham, S.~Jaitly and A.J.~Tolley, \emph{{QED positivity
  bounds}}, \href{https://doi.org/10.1103/PhysRevD.103.125020}{\emph{Phys. Rev.
  D} {\bfseries 103} (2021) 125020}
  [\href{https://arxiv.org/abs/2012.05798}{{\ttfamily 2012.05798}}].

\bibitem{Arkani-Hamed:2020blm}
N.~Arkani-Hamed, T.-C.~Huang and Y.-T.~Huang, \emph{{The EFT-Hedron}},
  \href{https://doi.org/10.1007/JHEP05(2021)259}{\emph{JHEP} {\bfseries 05}
  (2021) 259} [\href{https://arxiv.org/abs/2012.15849}{{\ttfamily
  2012.15849}}].

\bibitem{Caron-Huot:2021rmr}
S.~Caron-Huot, D.~Mazac, L.~Rastelli and D.~Simmons-Duffin, \emph{{Sharp
  Boundaries for the Swampland}},
  \href{https://doi.org/10.1007/jhep07(2021)110}{\emph{JHEP} {\bfseries 07}
  (2021) 110} [\href{https://arxiv.org/abs/2102.08951}{{\ttfamily
  2102.08951}}].

\bibitem{Arkani-Hamed:2021ajd}
N.~Arkani-Hamed, Y.-t.~Huang, J.-Y.~Liu and G.N.~Remmen, \emph{{Causality,
  Unitarity, and the Weak Gravity Conjecture}},
  \href{https://arxiv.org/abs/2109.13937}{{\ttfamily 2109.13937}}.

\bibitem{Chowdhury:2021ynh}
S.D.~Chowdhury, K.~Ghosh, P.~Haldar, P.~Raman and A.~Sinha, \emph{{Crossing
  Symmetric Spinning S-matrix Bootstrap: EFT bounds}},
  \href{https://arxiv.org/abs/2112.11755}{{\ttfamily 2112.11755}}.

\bibitem{Chiang:2022jep}
L.-Y.~Chiang, Y.-t.~Huang, W.~Li, L.~Rodina and H.-C.~Weng,
  \emph{{(Non)-projective bounds on gravitational EFT}},
  \href{https://arxiv.org/abs/2201.07177}{{\ttfamily 2201.07177}}.

\bibitem{Gell-Mann:1954ttj}
M.~Gell-Mann, M.L.~Goldberger and W.E.~Thirring, \emph{{Use of causality
  conditions in quantum theory}},
  \href{https://doi.org/10.1103/PhysRev.95.1612}{\emph{Phys. Rev.} {\bfseries
  95} (1954) 1612}.

\bibitem{Adams:2006sv}
A.~Adams, N.~Arkani-Hamed, S.~Dubovsky, A.~Nicolis and R.~Rattazzi,
  \emph{{Causality, analyticity and an IR obstruction to UV completion}},
  \href{https://doi.org/10.1088/1126-6708/2006/10/014}{\emph{JHEP} {\bfseries
  10} (2006) 014} [\href{https://arxiv.org/abs/hep-th/0602178}{{\ttfamily
  hep-th/0602178}}].

\bibitem{Chandrasekaran:2018qmx}
V.~Chandrasekaran, G.N.~Remmen and A.~Shahbazi-Moghaddam, \emph{{Higher-Point
  Positivity}}, \href{https://doi.org/10.1007/JHEP11(2018)015}{\emph{JHEP}
  {\bfseries 11} (2018) 015}
  [\href{https://arxiv.org/abs/1804.03153}{{\ttfamily 1804.03153}}].

\bibitem{Endlich:2017tqa}
S.~Endlich, V.~Gorbenko, J.~Huang and L.~Senatore, \emph{{An effective
  formalism for testing extensions to General Relativity with gravitational
  waves}}, \href{https://doi.org/10.1007/JHEP09(2017)122}{\emph{JHEP}
  {\bfseries 09} (2017) 122}
  [\href{https://arxiv.org/abs/1704.01590}{{\ttfamily 1704.01590}}].

\bibitem{Cardoso:2018ptl}
V.~Cardoso, M.~Kimura, A.~Maselli and L.~Senatore, \emph{{Black Holes in an
  Effective Field Theory Extension of General Relativity}},
  \href{https://doi.org/10.1103/PhysRevLett.121.251105}{\emph{Phys. Rev. Lett.}
  {\bfseries 121} (2018) 251105}
  [\href{https://arxiv.org/abs/1808.08962}{{\ttfamily 1808.08962}}].

\bibitem{Sennett:2019bpc}
N.~Sennett, R.~Brito, A.~Buonanno, V.~Gorbenko and L.~Senatore,
  \emph{{Gravitational-Wave Constraints on an Effective Field-Theory Extension
  of General Relativity}},
  \href{https://doi.org/10.1103/PhysRevD.102.044056}{\emph{Phys. Rev. D}
  {\bfseries 102} (2020) 044056}
  [\href{https://arxiv.org/abs/1912.09917}{{\ttfamily 1912.09917}}].

\bibitem{Ruhdorfer:2019qmk}
M.~Ruhdorfer, J.~Serra and A.~Weiler, \emph{{Effective Field Theory of Gravity
  to All Orders}}, \href{https://doi.org/10.1007/JHEP05(2020)083}{\emph{JHEP}
  {\bfseries 05} (2020) 083}
  [\href{https://arxiv.org/abs/1908.08050}{{\ttfamily 1908.08050}}].

\bibitem{Henriksson:2022oeu}
J.~Henriksson, B.~McPeak, F.~Russo and A.~Vichi, \emph{{Bounding Violations of
  the Weak Gravity Conjecture}},
  \href{https://arxiv.org/abs/2203.08164}{{\ttfamily 2203.08164}}.

\bibitem{Caron-Huot:2020cmc}
S.~Caron-Huot and V.~Van~Duong, \emph{{Extremal Effective Field Theories}},
  \href{https://doi.org/10.1007/JHEP05(2021)280}{\emph{JHEP} {\bfseries 05}
  (2021) 280} [\href{https://arxiv.org/abs/2011.02957}{{\ttfamily
  2011.02957}}].

\bibitem{Du:2021byy}
Z.-Z.~Du, C.~Zhang and S.-Y.~Zhou, \emph{{Triple crossing positivity bounds for
  multi-field theories}},
  \href{https://doi.org/10.1007/JHEP12(2021)115}{\emph{JHEP} {\bfseries 12}
  (2021) 115} [\href{https://arxiv.org/abs/2111.01169}{{\ttfamily
  2111.01169}}].

\bibitem{Vecchi:2007na}
L.~Vecchi, \emph{{Causal versus analytic constraints on anomalous quartic gauge
  couplings}}, \href{https://doi.org/10.1088/1126-6708/2007/11/054}{\emph{JHEP}
  {\bfseries 11} (2007) 054} [\href{https://arxiv.org/abs/0704.1900}{{\ttfamily
  0704.1900}}].

\bibitem{Bellazzini:2014waa}
B.~Bellazzini, L.~Martucci and R.~Torre, \emph{{Symmetries, Sum Rules and
  Constraints on Effective Field Theories}},
  \href{https://doi.org/10.1007/JHEP09(2014)100}{\emph{JHEP} {\bfseries 09}
  (2014) 100} [\href{https://arxiv.org/abs/1405.2960}{{\ttfamily 1405.2960}}].

\bibitem{Bellazzini:2016xrt}
B.~Bellazzini, \emph{{Softness and amplitudes\textquoteright{} positivity for
  spinning particles}},
  \href{https://doi.org/10.1007/JHEP02(2017)034}{\emph{JHEP} {\bfseries 02}
  (2017) 034} [\href{https://arxiv.org/abs/1605.06111}{{\ttfamily
  1605.06111}}].

\bibitem{deRham:2017avq}
C.~de~Rham, S.~Melville, A.J.~Tolley and S.-Y.~Zhou, \emph{{Positivity bounds
  for scalar field theories}},
  \href{https://doi.org/10.1103/PhysRevD.96.081702}{\emph{Phys. Rev. D}
  {\bfseries 96} (2017) 081702}
  [\href{https://arxiv.org/abs/1702.06134}{{\ttfamily 1702.06134}}].

\bibitem{deRham:2017zjm}
C.~de~Rham, S.~Melville, A.J.~Tolley and S.-Y.~Zhou, \emph{{UV complete me:
  Positivity Bounds for Particles with Spin}},
  \href{https://doi.org/10.1007/JHEP03(2018)011}{\emph{JHEP} {\bfseries 03}
  (2018) 011} [\href{https://arxiv.org/abs/1706.02712}{{\ttfamily
  1706.02712}}].

\bibitem{Tolley:2020gtv}
A.J.~Tolley, Z.-Y.~Wang and S.-Y.~Zhou, \emph{{New positivity bounds from full
  crossing symmetry}},
  \href{https://doi.org/10.1007/JHEP05(2021)255}{\emph{JHEP} {\bfseries 05}
  (2021) 255} [\href{https://arxiv.org/abs/2011.02400}{{\ttfamily
  2011.02400}}].

\bibitem{Bellazzini:2020cot}
B.~Bellazzini, J.~Elias~Mir\'o, R.~Rattazzi, M.~Riembau and F.~Riva,
  \emph{{Positive moments for scattering amplitudes}},
  \href{https://doi.org/10.1103/PhysRevD.104.036006}{\emph{Phys. Rev. D}
  {\bfseries 104} (2021) 036006}
  [\href{https://arxiv.org/abs/2011.00037}{{\ttfamily 2011.00037}}].

\bibitem{Chandorkar:2021viw}
D.~Chandorkar, S.D.~Chowdhury, S.~Kundu and S.~Minwalla, \emph{{Bounds on Regge
  growth of flat space scattering from bounds on chaos}},
  \href{https://doi.org/10.1007/JHEP05(2021)143}{\emph{JHEP} {\bfseries 05}
  (2021) 143} [\href{https://arxiv.org/abs/2102.03122}{{\ttfamily
  2102.03122}}].

\bibitem{Haring:2022cyf}
K.~H\"aring and A.~Zhiboedov, \emph{{Gravitational Regge bounds}},
  \href{https://arxiv.org/abs/2202.08280}{{\ttfamily 2202.08280}}.

\bibitem{Nicolis:2009qm}
A.~Nicolis, R.~Rattazzi and E.~Trincherini, \emph{{Energy's and amplitudes'
  positivity}}, \href{https://doi.org/10.1007/JHEP05(2010)095}{\emph{JHEP}
  {\bfseries 05} (2010) 095} [\href{https://arxiv.org/abs/0912.4258}{{\ttfamily
  0912.4258}}].

\bibitem{deRham:2017imi}
C.~de~Rham, S.~Melville, A.J.~Tolley and S.-Y.~Zhou, \emph{{Massive Galileon
  Positivity Bounds}},
  \href{https://doi.org/10.1007/JHEP09(2017)072}{\emph{JHEP} {\bfseries 09}
  (2017) 072} [\href{https://arxiv.org/abs/1702.08577}{{\ttfamily
  1702.08577}}].

\bibitem{Bellazzini:2017fep}
B.~Bellazzini, F.~Riva, J.~Serra and F.~Sgarlata, \emph{{Beyond Positivity
  Bounds and the Fate of Massive Gravity}},
  \href{https://doi.org/10.1103/PhysRevLett.120.161101}{\emph{Phys. Rev. Lett.}
  {\bfseries 120} (2018) 161101}
  [\href{https://arxiv.org/abs/1710.02539}{{\ttfamily 1710.02539}}].

\bibitem{deRham:2017xox}
C.~de~Rham, S.~Melville and A.J.~Tolley, \emph{{Improved Positivity Bounds and
  Massive Gravity}}, \href{https://doi.org/10.1007/JHEP04(2018)083}{\emph{JHEP}
  {\bfseries 04} (2018) 083}
  [\href{https://arxiv.org/abs/1710.09611}{{\ttfamily 1710.09611}}].

\bibitem{Baratella:2021guc}
P.~Baratella, D.~Haslehner, M.~Ruhdorfer, J.~Serra and A.~Weiler, \emph{{RG of
  GR from On-shell Amplitudes}},
  \href{https://arxiv.org/abs/2109.06191}{{\ttfamily 2109.06191}}.

\bibitem{Mignemi:1992nt}
S.~Mignemi and N.R.~Stewart, \emph{{Charged black holes in effective string
  theory}}, \href{https://doi.org/10.1103/PhysRevD.47.5259}{\emph{Phys. Rev. D}
  {\bfseries 47} (1993) 5259}
  [\href{https://arxiv.org/abs/hep-th/9212146}{{\ttfamily hep-th/9212146}}].

\bibitem{Noller:2019chl}
J.~Noller, L.~Santoni, E.~Trincherini and L.G.~Trombetta, \emph{{Black Hole
  Ringdown as a Probe for Dark Energy}},
  \href{https://doi.org/10.1103/PhysRevD.101.084049}{\emph{Phys. Rev. D}
  {\bfseries 101} (2020) 084049}
  [\href{https://arxiv.org/abs/1911.11671}{{\ttfamily 1911.11671}}].

\bibitem{Lyu:2022gdr}
Z.~Lyu, N.~Jiang and K.~Yagi, \emph{{Constraints on
  Einstein-dilation-Gauss-Bonnet gravity from black hole-neutron star
  gravitational wave events}},
  \href{https://doi.org/10.1103/PhysRevD.105.064001}{\emph{Phys. Rev. D}
  {\bfseries 105} (2022) 064001}
  [\href{https://arxiv.org/abs/2201.02543}{{\ttfamily 2201.02543}}].

\bibitem{deRham:2020ejn}
C.~de~Rham, J.~Francfort and J.~Zhang, \emph{{Black Hole Gravitational Waves in
  the Effective Field Theory of Gravity}},
  \href{https://doi.org/10.1103/PhysRevD.102.024079}{\emph{Phys. Rev. D}
  {\bfseries 102} (2020) 024079}
  [\href{https://arxiv.org/abs/2005.13923}{{\ttfamily 2005.13923}}].

\bibitem{AccettulliHuber:2020dal}
M.~Accettulli~Huber, A.~Brandhuber, S.~De~Angelis and G.~Travaglini,
  \emph{{From amplitudes to gravitational radiation with cubic interactions and
  tidal effects}},
  \href{https://doi.org/10.1103/PhysRevD.103.045015}{\emph{Phys. Rev. D}
  {\bfseries 103} (2021) 045015}
  [\href{https://arxiv.org/abs/2012.06548}{{\ttfamily 2012.06548}}].

\bibitem{Cano:2019ore}
P.A.~Cano and A.~Ruip\'erez, \emph{{Leading higher-derivative corrections to
  Kerr geometry}}, \href{https://doi.org/10.1007/JHEP05(2019)189}{\emph{JHEP}
  {\bfseries 05} (2019) 189}
  [\href{https://arxiv.org/abs/1901.01315}{{\ttfamily 1901.01315}}].

\bibitem{Cano:2021myl}
P.A.~Cano, K.~Fransen, T.~Hertog and S.~Maenaut, \emph{{Gravitational ringing
  of rotating black holes in higher-derivative gravity}},
  \href{https://doi.org/10.1103/PhysRevD.105.024064}{\emph{Phys. Rev. D}
  {\bfseries 105} (2022) 024064}
  [\href{https://arxiv.org/abs/2110.11378}{{\ttfamily 2110.11378}}].

\bibitem{Silva:2022srr}
H.O.~Silva, A.~Ghosh and A.~Buonanno, \emph{{Black-hole ringdown as a probe of
  higher-curvature gravity theories}},
  \href{https://arxiv.org/abs/2205.05132}{{\ttfamily 2205.05132}}.

\bibitem{Nair:2019iur}
R.~Nair, S.~Perkins, H.O.~Silva and N.~Yunes, \emph{{Fundamental Physics
  Implications for Higher-Curvature Theories from Binary Black Hole Signals in
  the LIGO-Virgo Catalog GWTC-1}},
  \href{https://doi.org/10.1103/PhysRevLett.123.191101}{\emph{Phys. Rev. Lett.}
  {\bfseries 123} (2019) 191101}
  [\href{https://arxiv.org/abs/1905.00870}{{\ttfamily 1905.00870}}].

\bibitem{Silva:2020acr}
H.O.~Silva, A.M.~Holgado, A.~C\'ardenas-Avenda\~no and N.~Yunes,
  \emph{{Astrophysical and theoretical physics implications from multimessenger
  neutron star observations}},
  \href{https://doi.org/10.1103/PhysRevLett.126.181101}{\emph{Phys. Rev. Lett.}
  {\bfseries 126} (2021) 181101}
  [\href{https://arxiv.org/abs/2004.01253}{{\ttfamily 2004.01253}}].

\bibitem{Cano:2020cao}
P.A.~Cano, K.~Fransen and T.~Hertog, \emph{{Ringing of rotating black holes in
  higher-derivative gravity}},
  \href{https://doi.org/10.1103/PhysRevD.102.044047}{\emph{Phys. Rev. D}
  {\bfseries 102} (2020) 044047}
  [\href{https://arxiv.org/abs/2005.03671}{{\ttfamily 2005.03671}}].

\bibitem{Wagle:2021tam}
P.~Wagle, N.~Yunes and H.O.~Silva, \emph{{Quasinormal modes of slowly-rotating
  black holes in dynamical Chern-Simons gravity}},
  \href{https://arxiv.org/abs/2103.09913}{{\ttfamily 2103.09913}}.

\bibitem{Srivastava:2021imr}
M.~Srivastava, Y.~Chen and S.~Shankaranarayanan, \emph{{Analytical computation
  of quasinormal modes of slowly rotating black holes in dynamical Chern-Simons
  gravity}}, \href{https://doi.org/10.1103/PhysRevD.104.064034}{\emph{Phys.
  Rev. D} {\bfseries 104} (2021) 064034}
  [\href{https://arxiv.org/abs/2106.06209}{{\ttfamily 2106.06209}}].

\bibitem{Dima:2020yac}
A.~Dima, E.~Barausse, N.~Franchini and T.P.~Sotiriou, \emph{{Spin-induced black
  hole spontaneous scalarization}},
  \href{https://doi.org/10.1103/PhysRevLett.125.231101}{\emph{Phys. Rev. Lett.}
  {\bfseries 125} (2020) 231101}
  [\href{https://arxiv.org/abs/2006.03095}{{\ttfamily 2006.03095}}].

\end{thebibliography}\endgroup

\end{document}